\documentclass{emulateapj}

\usepackage{graphicx}
\usepackage{subfigure}
\usepackage[hypertex]{hyperref}

\def \th {\thinspace}
\def \nh {N${\rm _H}$}

\def \arcmin {\hbox{$^\prime$}}
\def \arcsec {\hbox{$^{\prime\prime}$}}

\def\spose#1{\hbox to 0pt{#1\hss}}
\def\ltsim{$\mathrel{\spose{\lower 3pt\hbox{$\sim$}}
        \raise 2.0pt\hbox{$<$}}$\thinspace}
\def\gtsim{$\mathrel{\spose{\lower 3pt\hbox{$\sim$}}
        \raise 2.0pt\hbox{$>$}}$\thinspace}
\def \msun {${\rm M_\odot}$}

\def \nh {$N_{\rm H}$}

\def \eg {e.g.}

\def \ie {i.e.}

\def \dtwentyfive {${\rm D_{25}}$}
\newcommand{\first}{${\rm 1^{st}}$}
\newcommand{\second}{${\rm 2^{nd}}$}
\newcommand{\third}{${\rm 3^{rd}}$}
\newcommand{\fourth}{${\rm 4^{th}}$}
\newcommand{\fifth}{${\rm 5^{th}}$}
\newcommand{\sixth}{${\rm 6^{th}}$}

\newcommand{\apec}{APEC}
\newcommand{\mekal}{MEKAL}
\newcommand{\hnought}{${\rm H_0}$}
\newcommand{\zfe }{${\rm Z_{Fe}}$}
\newcommand{\zo }{${\rm Z_{O}}$}
\newcommand{\zsi }{${\rm Z_{Si}}$}
\newcommand{\zni }{${\rm Z_{Ni}}$}
\newcommand{\zs }{${\rm Z_{S}}$}
\newcommand{\zne }{${\rm Z_{Ne}}$}
\newcommand{\zmg }{${\rm Z_{Mg}}$}
\newcommand{\feh } {${\rm [Fe/H]}$}
\newcommand{\chandra }{\href{http://chandra.harvard.edu/}{\em Chandra}}
\newcommand{\xspec }{\href{http://heasarc.gsfc.nasa.gov/docs/software/lheasoft/xanadu/xspec/index.html}{\em Xspec}}
\newcommand{\acis }{{\em ACIS}}
\newcommand{\ciao }{\href{http://cxc.harvard.edu/ciao/}{\em CIAO}}
\newcommand{\caldb }{\href{http://cxc.harvard.edu/caldb/}{\em Caldb}}
\newcommand{\heasoft }{\href{http://heasarc.gsfc.nasa.gov/docs/software/lheasoft/}{\em Heasoft}}
\newcommand{\ergps}{${\rm erg\ s^{-1}}$}
\newcommand{\ergpscm}{${\rm erg\ s^{-1} cm^{-2}}$}
\newcommand{\xmm }{{\em XMM}}
\newcommand{\asca }{{\em ASCA}}
\newcommand{\rosat }{{\em Rosat}}

\newcommand{\lx }{${\rm L_X}$}
\newcommand{\lb }{${\rm L_B}$}
\newcommand{\fgas}{${\rm f^{Ia}_{gas}}$}

\newcommand{\fstars}{${\rm f^{Ia}_{*}}$}
\newcommand{\lsun }{${\rm L_\odot}$}
\newcommand{\leda}{{\em{LEDA}}}
\newcommand{\ned}{{\em{NED}}}

\newcommand{\thin}{\thinspace}
\slugcomment{Accepted for publication in the Astrophysical Journal}
\shorttitle{ISM abundances in early-type galaxies.}
\shortauthors{Humphrey \& Buote.}

\begin{document}

\title{A Chandra Survey of Early-type Galaxies, I: Metal Enrichment in the ISM.}
\author {\href{mailto:phumphre@uci.edu}{Philip J. Humphrey}\altaffilmark{1} and  David A. Buote\altaffilmark{1}}
\altaffiltext{1}{Department of Physics and Astronomy, University of California at Irvine, 4129 Frederick Reines Hall, Irvine, CA 92697-4575}
\begin{abstract}
We present the first in a series of papers studying with \chandra\
the X-ray properties of a sample of 28 early-type galaxies which span
$\sim$3 orders of magnitude in X-ray luminosity (\lx). We report
emission-weighted Fe abundance (\zfe) constraints and, for many of the 
galaxies,
abundance constraints for  key elements such as O, Ne, Mg, Si, S and Ni. 
We find no evidence of the very sub-solar \zfe\ historically reported,
confirming a trend in recent X-ray observations of bright galaxies
and groups, nor do we find any correlation between \zfe\ and luminosity.
Except in one case we do not find evidence for a multi-phase interstellar
medium (ISM), indicating that multi-temperature fits required in previous
\asca\ analysis arose due to the strong temperature gradients which
we are able to resolve with \chandra. We compare the stellar \zfe, estimated
from simple stellar population model fits, to that of the hot gas.  
Excepting one possible outlier we find no evidence that the gas is substantially
more metal-poor than the stars and, in a few systems, \zfe\ is higher
in the ISM. In general, however, the two components exhibit 
similar metallicities, which is inconsistent with both 
galactic wind models and recent hierarchical chemical enrichment
simulations. Adopting standard SNIa and SNII metal yields our abundance ratio 
constraints imply 66$\pm$11\% of the Fe within the ISM was produced
in SNIa, which is remarkably similar to the Solar neighbourhood, and implies
similar enrichment histories for the cold ISM in a spiral and the hot
ISM in elliptical galaxies. Although these values are sensitive to the
considerable systematic uncertainty in the supernova yields, they are 
also in very good agreement with observations of more massive systems.
These results indicate a remarkable degree of homology in the enrichment
process operating from cluster scales to low-to-intermediate \lx\ galaxies.
In addition the data uniformly exhibit the low \zo/\zmg\ abundance ratios which
have been reported in the centres of clusters, groups and some galaxies.
This is inconsistent with the standard calculations of metal production in 
SNII and may indicate an additional source of 
$\alpha$-element enrichment, such as Population~III hypernovae. 
\end{abstract}


\section{Introduction}
The entire history of star-formation and evolution leaves its
chemical signature in the hot gas  of early-type galaxies. 
X-ray observations therefore provide a natural and powerful diagnostic tool
to unlock this information \citep[\eg][]{loewenstein91,mathews03a}. However,
historically X-ray measurements of interstellar medium (ISM) abundances have 
been
problematical, as typified by the so-called ``Fe discrepancy''
\citep{arimoto97}. Early \rosat\ and \asca\ observations of these
galaxies tended to imply extremely sub-solar metal abundances (generally
expressed as \zfe, the Fe abundance with respect to the adopted
solar standard, since Fe has the strongest diagnostic 
lines in the soft X-ray band) \citep[\eg][]{loewenstein98}, in stark 
contrast to $\sim$solar abundances in the stellar population.
Since individual galaxies are not ``closed boxes'', the ISM is believed
to be built up primarily through stellar mass-loss and 
Type Ia supernovae (SNe) ejecta. These two crucial ingredients lead classical 
``wind models'' of gas enrichment 
to predict highly super-solar \zfe\ in these galaxies
\citep[\eg][]{ciotti91,loewenstein91}. The problems caused by this
discrepancy are  exacerbated further when attempting to understand
gas enrichment in clusters of galaxies, for which X-ray observations
typically find \zfe$\sim$0.3--0.5. The metal content of the intra-cluster
medium (ICM) is primarily attributed to the stellar ejecta from 
giant elliptical galaxies, and so it is difficult to envisage
there being {\em lower} metal abundances in individual galaxies than
in the ICM \citep{renzini97}.

Attempts to reproduce the low \zfe\ in early-type galaxies from a modelling 
standpoint by, for example, incorporating ongoing accretion of unenriched
gas \citep[\eg][]{brighenti99a} or allowing complex star-formation histories
\citep[\eg][]{kawata03a} have not been entirely successful, leading some
to point a finger at the spectral-modelling. \citet{arimoto97},
for instance, called into question the plasma codes being fitted to
the data, particularly in light of uncertainties associated with the 
Fe L-shell transitions in X-ray emission plasma codes. 
This point was investigated further by
\citet{matsushita00a}, who argued that, having degraded the data quality
in the vicinity of the Fe L-shell, \zfe \gtsim0.5 can be 
found in the highest X-ray luminosity (\lx) galaxies. Consistent
results from spectral-fitting with  plasma codes that treat the
Fe L-shell transitions differently would seem, however,
to conflict with this explanation \citep[\eg][]{buote03b}. 
Although remaining errors in the treatment of the Fe L-shell in the 
plasma codes  can be 
important for high-resolution spectroscopy \citep[\eg][]{behar01},
at CCD resolution the effects are substantially washed out so that 
they contribute only a $\sim$10--20\% systematic
uncertainty to the abundance measurement \citep{buote03b}. 

Perhaps
a more natural alternative was suggested by \citet{buote98c},
who demonstrated that fitting a single temperature model to 
the emission spectrum of intrinsically
non-isothermal hot gas gives rise to a substantially under-estimated 
abundance (not to mention a significantly poorer fit), an effect 
dubbed the ``Fe bias'' \citep[see also][]{buote00a}.  
This effect had previously been recognized by \citet{buote94}, and
\citet{trinchieri94} found that fitting a two-temperature model
to the \rosat\ spectrum of the bright elliptical NGC\thin 4636 resulted
in poorly-constrained abundances which could be consistent with solar
values, in contrast to low \zfe\ required for single-temperature fits.
A spectrally hard (kT\gtsim 5~keV) component had long been recognized
in the \asca\ spectra of the many early-type galaxies, which was attributed
to emission from undetected X-ray binaries \citep{matsushita94}.
The composite spectrum from these sources, which dominate the emission
of low-\lx\ galaxies, can typically be approximated as a pure bremsstrahlung 
model. Therefore, fitting only a single-temperature hot gas model to the 
spectra of such galaxies also tends to give rise to unphysically low 
abundances \citep[\eg][who found essentially unconstrained abundances
when incorporating a term to account for this effect in their \rosat\
analysis]{fabbiano94,kim96}. This effect is clearly distinct
from, albeit related to, the Fe bias \citep[see discussion in][]{buote00a}.
Although the spectral-shape of the unresolved source component 
is sufficiently hard that it cannot move to mitigate, even in part, 
the Fe bias arising from multi-temperature hot gas components, 
\citet{buote98c} still found \zfe\ consistent with solar in 
the lower-\lx\ systems, in which
the fit implied only  a single hot gas component, plus undetected
sources. However, the poor signal-to-noise ratio (S/N) characteristic
of the lower-\lx\ galaxies tended to produce poorly-constrained abundances,
so that very sub-solar abundances could not be ruled out.

Recent \chandra\ and \xmm\ measurements of abundances in X-ray bright
galaxies and the centres of groups have tended to support
$\sim$solar or even slightly super-solar abundances for the ISM
\citep[\eg][]{buote03b,tamura03a,gastaldello02a,buote02a,xu02a,osullivan03b,kim04a}. In contrast, very sub-solar \zfe\ are still being reported
in the lowest \lx/\lb\ systems \citep[\eg][]{irwin02a,sarazin01}. Perhaps
the most dramatic examples of this latter effect are the three very 
low-\lx\ galaxies for which \citet{osullivan04a} reported 
\zfe\ltsim 0.1. This apparent lack of consistency in the enrichment
processes operating on different scales is intriguing, and it remains
to be assessed whether it is an artefact of the poorer S/N or a real
effect in low-\lx\ systems. 
Although a major problem for our understanding of galaxy evolution,
a full understanding of this effect is inhibited by the lack of 
interesting abundance constraints in galaxies with \lx\ intermediate between these
two extremes. In a recent paper, \citet{humphrey04b}, 
we made some initial progress in this area by reporting constraints for
the normal, moderate-\lx\ S0 galaxy NGC\thin 1332 and the elliptical
NGC\thin 720, in both cases strongly excluding the extremely
sub-solar (\zfe \ltsim 0.4) abundances historically reported for
these systems. Coupled with similar
constraints on the abundances in the moderate \lx/\lb\ radio galaxy
NGC\thin 1316 \citep{kim03b}, this would hint at  a consistent
picture of enrichment from cluster to moderate-\lx\ galaxy
scales.

In addition to the insight into enrichment afforded from global
\zfe\ measurements, the $\alpha$-element abundance ratios, with 
respect to Fe, provide an additional powerful diagnostic. Different
types of SNe inject material imprinted with characteristic chemical
``fingerprints'', so that the measured abundance pattern can be
used to assess the relative contribution of SNIa and SNII to ISM
enrichment. Early attempts to extract this information 
\citep[\eg][]{matsushita00a,finoguenov00a} were largely affected by 
a failure to treat the Fe bias, although a significant contribution
of SNIa to the enrichment process was indicated. More recently, studies
with high-quality \xmm\ and \chandra\ data have revealed typically
$\sim$70--90\% of the ISM enrichment in the centres of groups and
clusters seems to have its origin in SNIa, which is close to 
the $\sim$75\% value in the Solar neighbourhood 
\citep[\eg][]{gastaldello02a,buote03b}. For NGC\thin 1332 and NGC\thin 720,
we were able to obtain interesting constraints on the $\alpha$-to-Fe ratios
in some of the lowest-\lx\ systems to date, again inferring $\sim$70--80\%
of the Fe to have been produced in SNIa. This also supports the suggestion
of homology in the enrichment process over different mass-scales, at least
down to moderate-\lx\ galaxies. 

Although these results are intriguing, it is by no means clear that 
the two galaxies we considered are representative, nor is it clear 
that all X-ray bright galaxies follow the trend of $\sim$solar
abundances \citep[\eg][]{sambruna04a}. Furthermore,
these studies have not elucidated the processes which may 
be giving rise to very low-\zfe\ measurements in the faintest systems. 
In light of these results, therefore, the next logical step is to consider a
uniformly-analyzed sample of galaxies spanning a large range of \lx,
and particularly expanding the number of moderate-\lx\ galaxies studied.
In this paper, we present the results of a study of metal enrichment 
in a sample of 28 early-type galaxies drawn from the \chandra\ archive.
In subsequent papers we will discuss the gravitating mass and point-source
populations of these objects as a whole.
The galaxies have been carefully selected to span the available X-ray 
luminosity range, from group-dominant to low-\lx\ galaxies.

In many respects, \chandra\ ACIS is the natural instrument with which to
undertake such a study. Although \xmm\ has a significantly higher 
collecting area, the analysis of faint, diffuse sources is complicated by 
difficulties in treating the background. \chandra\ also has an intrinsic
advantage in being able to resolve out a substantial fraction of the 
X-ray binary contribution into individual sources, which must otherwise
be disentangled spectrally. 
The excellent spatial resolution of \chandra\ also provides an unprecedented 
opportunity to investigate any spatial temperature variation, which would 
provide the natural source of the Fe bias. 
Imaging spectroscopy at CCD resolution is well-suited to determine
reliable abundances in a galaxy. In fact there are a number of 
drawbacks to using the \chandra\ and \xmm\ gratings instead for such a study.
The extended nature of the sources makes grating spectroscopy 
extremely challenging. The significantly lower effective area
of the grating spectrographs in comparison to the ACIS CCDs and,
especially for the \xmm\ RGS instrument, the more limited bandwidth
both exacerbate this problem. Although one of the key issues of 
interest is the determination of multiple temperature components
in the extraction aperture, this tends to produce features broad
enough to be evident in CCD spectra. In fact, based on high S/N data
of the group NGC\thin 5044 there was excellent agreement in the 
abundances determined with the \xmm\ gratings and the
\xmm\ and \chandra\ CCDS \citep{buote03b,tamura03a}, confirming that
CCD resolution is sufficient to obtain reliable abundances.

Throughout this paper we adopt the latest solar abundances standard of 
\citet[][see \S~\ref{sect_solar_standard}]{asplund04a}, which resolve
several previously-noted discrepancies between ``photospheric'' and 
``meteoritic'' values. All error-bars quoted
refer to the 90\% confidence region, unless otherwise stated.

\section{Target Selection}
By its nature, the \chandra\ archive contains a nonuniform sample 
of galaxies.
Although this prevents our choosing a statistically complete sample,
we selected 28 galaxies
approximately spanning the range of measured \lx, from 
$\sim 8\times 10^{39}$--$10^{43}$\ergps.
We only considered relatively nearby galaxies which had been observed for a 
total of at least 10~ks
with the ACIS-S or ACIS-I, and without any grating employed.
For selection purposes X-ray luminosities were taken
from the catalogue of \citet{osullivan01a}, although all 
luminosities have subsequently been recomputed in the present work 
(\S~\ref{sect_lx}). To an initial sample of 26 galaxies chosen in
this way, we also
added two further interesting objects not in the O'Sullivan catalogue---
NGC\thin 1132, one of the closest examples of a ``fossil group'' 
\citep[][F. Gastaldello et al, 2005, in prep.]{mulchaey99a} and 
the relatively isolated, moderate-\lx\ galaxy NGC\thin 1700 which
has an unusually high X-ray ellipticity, which \citet{statler02a}
argued may indicate rotational support.
Two of our sample, NGC\thin 1332 and NGC\thin 720, have already
been discussed in \citet{humphrey04b}, and we adopt the results
from that work here.
The sample 
contains 10 high-\lx\ (${\rm log_{10}}$\lx\gtsim 41.5) galaxies,  
12 moderate-\lx (${\rm log_{10}}$\lx$\simeq$40.5--41.5)
and 6 low-\lx\ (${\rm log_{10}}$\lx\ltsim 40.5 ) galaxies.
Most of the sample was observed with ACIS-S, although in a few 
cases ACIS-I was employed. To give extra coverage at large
radii both ACIS-S and ACIS-I data for the two bright
galaxies NGC\thin 1399 and NGC\thin 4472, were analysed together.

A summary of the properties of the galaxies and details of the 
\chandra\ exposures are given in Table~\ref{table_obs}.
In order to provide accurate luminosity estimates, we searched the 
literature for reliable distance estimates. Where possible, we 
adopted those determined from surface brightness fluctuations (SBF) by 
\citet[][correcting for an improved Cepheid zero-point: \citealt{jensen03}]{tonry01}
or \citet{jensen03}. Alternatively, we used distances determined from the 
${\rm D_n-\sigma}$ relation  \citep{faber89}, or the redshift, corrected 
for Virgo-centric flow, as given in \leda. We assumed 
\hnought=70 ${\rm km\ s^{-1}\ Mpc^{-1}}$.

Of this sample, 6 systems (IC\thin 4296, NGC\thin 507, NGC\thin 741,
NGC\thin 1399, NGC\thin 4472, NGC\thin 7619) appear to be the central 
galaxies in substantial, optically-identified
groups so their X-ray emission may be to some extent intertwined with
that of a surrounding intra-group medium (IGM). 
In the present context, it suffices to consider that all the systems 
comprise a continuum over a range of mass-scales. In a subsequent
paper, we will discuss the
issue of group membership, and the total gravitating mass, in detail.

%\clearpage
\renewcommand{\tabcolsep}{0.5mm}
\begin{deluxetable*}{llrrrrrrrrr}
\tablecaption{The sample\label{table_obs}}
\tabletypesize{\scriptsize}
\tablehead{
\colhead{Galaxy} & \colhead{Type} & \colhead{Dist} & \colhead{\dtwentyfive} & 
\colhead{\lb} & \colhead{log$_{10}$\lx} &  \colhead{\nh} & \colhead{ObsID} & \colhead{Instr.}& \colhead{Date} & 
\colhead{Exposure} \\
\colhead{} & \colhead{} & \colhead{(Mpc)} & \colhead{(\arcmin)} & 
\colhead{($10^{10}$\lsun)} & \colhead{(${\rm log_{10}(erg\ s^{-1})}$)} & \colhead{(${\rm 10^{20} cm^{-2}}$)} & \colhead{} & \colhead{} & 
\colhead{(dd/mm/yy)} & \colhead{(ks)} 
}
\startdata
\multicolumn{11}{l}{High-\lx\ galaxies} \\ \hline
IC 4296& E Radio gal      & 50.8$^2$       & 3.8&12.1& 41.54  &4.1 &3394     &S&10/09/01         &25\\
NGC 507&  SA(r)0     & 82.6$^3$       & 3.2&14.9& 43.03       &5.4 &2882     &I&08/01/02         &43\\
NGC 741 & E0             & 75.8$^3$       & 2.9&14.2& 42.32   &4.4 &2223     &S&28/01/01         &30\\
NGC 1132& E               & 98.2$^4$       & 2.1&8.3 & 42.76  &5.2 &801      &S&10/12/99         &13\\
NGC 1399& cD;E1 pec       & 18.5$^1$       & 6.9&4.2 & 42.09  &1.3 &319      &S&18/01/00         &56\\
        &                 &                &    &    &        &    &4174     &I&28/05/03         &45\\
NGC 1600& E3              & 57.4$^3$       & 3.2&9.8 & 42.19  &4.8 &4283     &S&18/09/02         &22\\
NGC 4472& E2/S0(2) Sy2    & 15.1$^1$       & 9.7&7.5 & 41.52  &1.7 &321      &S&12/06/00         &32\\
        &                 &                &    &    &        &    &322      &I&19/03/00         &10\\
NGC 5846& E0-1;LINER HII  & 21.1$^1$       & 3.8&3.1 & 41.57  &4.3 &788      &S&24/05/00         &23\\
NGC 7619& E               & 49.2$^1$       & 2.6&6.9 & 42.06  &5.0 &3955     &S&24/09/03         &31\\ 
NGC 7626& E pec          & 51.2$^3$       & 2.7&6.9 & 41.51   &5.0 &2074     &I&20/08/01         &26\\
\hline \multicolumn{11}{l}{Moderate-\lx\ galaxies} \\ \hline
NGC 720 & E5              & 25.7$^1$       & 4.6&3.1 & 41.33  &1.5 &492      &S&12/10/00         &29\\
NGC 1332& S(s)0      & 21.3$^1$       & 4.1&2.3 & 40.95       &2.2 &4372     &S&19/09/02         &45\\
NGC 1387& SAB(s)0        & 18.9$^1$       & 3.3&1.2 & 40.78   &1.3 &4168     &I&20/05/03         &45\\ 
NGC 1407& E0              & 26.8$^1$       & 5.3&6.5 & 41.32  &5.4 &791      &S&16/08/00         &40\\
NGC 1549& E0-1            & 18.3$^1$       & 4.7&2.9 & 40.66  &1.5 &2077     &S&08/11/00         &22\\
NGC 1553&SA(rl)0  LINER& 17.2$^1$       & 5.3&3.7 & 40.64     &1.5 &783      &S&02/01/00         &14\\
NGC 1700& E4              & 41.1$^1\dagger$& 3.0&4.6 & 41.20  &4.8 &2069     &S&03/11/00         &27\\
NGC 3607& SA(s)0     & 21.2$^1$       & 4.5&3.4 & 40.94       &1.5 &2073     &I&12/06/01         &38\\
NGC 3923& E4-5            & 21.3$^1$       & 6.4&4.9 & 41.03  &1.5 &1563     &S&14/06/01         &8.8\\
NGC 4365& E3              & 19.0$^1$       & 5.8&3.7 & 40.83  &6.2 &2015     &S&02/06/01         &40\\
NGC 4552& E;LINER HII     & 14.3$^1$       & 5.0&2.0 & 40.65  &2.6 &2072     &S&22/04/01         &54\\
NGC 5018& E3              & 42.6$^3$       & 3.6&7.1 & 40.96  &7.0 &2070     &S&14/04/01         &28\\
\hline \multicolumn{11}{l}{Low-\lx\ galaxies} \\ \hline
NGC 3115& S0             & 9.0$^1$        & 7.3&1.4 & 39.86   &4.3 &2040     &S&14/06/01         &36\\
NGC 3585& E7/S0           & 18.6$^1$       & 6.1&3.3 & 40.35  &5.6 &2078     &S&03/06/01         &35\\
NGC 3608& E2 LINER        & 21.3$^1$       & 3.2&1.7 & 40.39  &1.5 &2073     &I&12/06/01         &38\\
NGC 4494& E1-2 LINER      & 15.8$^1$       & 4.5&2.3 & 40.36  &1.5 &2079     &S&05/08/01         &15\\
NGC 4621& E5              & 17.0$^1$       & 5.0&2.5 & 40.14  &2.2 &2068     &S&01/08/01         &25\\
NGC 5845& E               & 24.0$^1$       & 0.9&0.45& 40.17  &4.3 &4009     &S&03/01/03         &30\\
\enddata
\tablecomments{Listed above are all of the galaxies in our sample.
Distances were obtained from $^1$--- SBF: \citet{tonry01}, corrected
for the the new Cepheid zero-point (see text), $^2$--- SBF: 
\citet{jensen03}, $^3$--- ${\rm D_n}$-$\sigma$: \citet{faber89},
$^4$--- redshift distance (\leda);
$\dagger$--- uncertain. 
\lb\ was determined from the face-on, reddening-corrected B-band magnitude
given by \leda. \lx\ was computed in the 0.1--10.0~keV band
self-consistently in the present work(\S~\ref{sect_lx}), excluding obvious 
emission from any low-luminosity AGN, and extrapolating the surface brightness 
to a  fiducial 300~kpc radius. The galaxy type was taken from \ned. 
\nh\ is the nominal Galactic column-density along the line-of-sight.
We show
the \chandra\ observation identifier (ObsID), the \acis\ instrument
(I or S) and the net exposure-time, having excluded periods of flaring.
}
\end{deluxetable*}

%\clearpage

\section{Data reduction} \label{sect_reduction}
For data reduction we used the \ciao\thin 3.1 and \heasoft\thin 5.3
software suites, in conjunction with \chandra\ \caldb\ calibration
database 2.28. For spectral-fitting we used \xspec\ 11.3.1.
In order to ensure the most up-to-date calibration, all data were
reprocessed from the ``level 1'' events files, following the 
standard \chandra\ data-reduction 
threads\footnote{\href{http://cxc.harvard.edu/ciao/threads/index.html}{http://cxc.harvard.edu/ciao/threads/index.html}}.
We applied corrections to take account of a time-dependent 
drift in the satellite gain and, for ACIS-I observations, the 
effects of ``charge transfer inefficiency'', as implemented in the 
standard \ciao\ tools.

To identify periods of enhanced background (``flaring''), which seriously
degrades the signal-to-noise (S/N) and complicates background subtraction
\citep[\eg][]{markevitch02},
we accumulated background lightcurves for each exposure from 
low surface-brightness regions of the active chips. We 
excluded obvious diffuse emission and data in the vicinity of any detected 
point-sources (see below). Periods of flaring were identified by eye and 
excised.
 Any residual flaring which is not removed by this procedure will be 
sufficiently mild to have negligible impact in the centres of bright
galaxies. However, in the fainter systems even very mild background
variation can have a significant impact on our results, and so we treat
these systems with extra care (\S~\ref{sect_systematics_bkg}).
The final exposure times are listed in Table~\ref{table_obs}. 

Point source detection was performed using the \ciao\ tool
{\tt wavdetect} \citep{freeman02}. In order to improve the likelihood of identifying sources
with peculiarly hard or soft spectra, full-resolution
images were created of the region of the
\acis\ focal-plane containing the S3 chip in the energy-band 
0.1--10.0~keV and, so as to identify any unusually soft
or hard sources, also in the bands 0.1--3.0~keV and 3.0--10.0~keV. Sources were
detected separately in each image. In order to minimize spurious detections at
node or chip boundaries we supplied the detection algorithm with
exposure-maps  generated at
energies 1.7~keV, 1.0~keV and 7~keV respectively (although the precise
energies chosen made little difference to the results). The
detection algorithm searched for structure over pixel-scales of 1, 2, 4, 8 and
16 pixels, and the detection threshold was set to $\sim 10^{-7}$ spurious
sources per pixel (corresponding to $\sim$0.1 spurious detections per
image). The source-lists obtained within each energy-band were combined and
duplicated sources removed, and the final list was checked
by visual inspection of the images. A full discussion of the 
point source populations will be given in a subsequent paper.
In the present work, the data in the vicinity of any detected point source
were removed so as not to contaminate the diffuse emission. 
As discussed in \citet[][see also \citealt{kim03a}]{humphrey04a}
a significant fraction of faint X-ray binary sources 
will not have been detected by this procedure, and so we include
an additional component to account for it in our spectral fitting.

\subsection{Background estimation} \label{sect_flare}
One of the key challenges in spectral-fitting
diffuse X-ray emission is ensuring proper background subtraction. 
For \chandra\ a set of blank-field event files
have been made available as part of the standard \caldb\ distribution,
from which background spectra can be accumulated corresponding to
similar regions of the detector. For each observation, we prepared from these
a suitably projected background events file. We were able to 
extract from this file ``template'' background spectra for each region of
the detector in which our ``source'' spectra were obtained.
However, these background spectra are unlikely to represent perfectly
the background in any one observation. There are known to be 
significant long-term secular variations in the non X-ray components
of the background, substantial field-to-field variation in the 
cosmic component, and there may be some residual mild flaring.
It is also worth noting that the hard (power law) component of the 
cosmic X-ray background arises from undetected background AGN, so
 its absolute normalization is also a strong function of the point
source detection completeness; in turn this is a function of the 
surface brightness of the galaxy and the total exposure time
\citep{kim03a}.

Several authors have adopted the practice of renormalizing the background
template to ensure good agreement with the instrumental background at high
energies (\gtsim 10~keV). Such a procedure, however, also 
renormalizes the (uncorrelated) cosmic X-ray background and instrumental 
line features, which can lead to serious over or under-subtraction. 
Given these reservations we chose to use an alternative 
background estimation procedure. 
Our  method involved modelling the background, somewhat akin to
the approach of  \citet{buote04c}. For each observation, we extracted
a spectrum from a ``source free'' region of the ACIS field of view.
We chose a $\sim$2\arcmin\ region centred on the S1 chip if the 
galaxy was centred on S3, or on the S2 chip where the galaxy 
was centred on ACIS-I. If the S1 or S2 chips were turned off, we 
chose a small, \ltsim 1\arcmin\ region on the S3 or ACIS-I chips, as 
appropriate, positioned to be in a region of as low surface-brightness
as possible.
We excluded data from the vicinity
of any point-sources found by the source detection algorithm.
Additionally, we extracted a ``source+background'' spectrum from
the CCD on which the source was centred, in an annulus 
centred at the galaxy 
centroid and with an inner and outer radii typically $\sim$2.5\arcmin\
and 3.3\arcmin.
We adopted two spectra since we found that this procedure enabled us 
most cleanly to constrain the background. Some of the brightest galaxies
are so extended in the X-ray that even in our ``source free'' region there
is a small contribution from hot gas at large radii. We found that 
using two spectra with different hot gas contributions allowed this 
to be readily disentangled from the actual background components.
In order to constrain the model, we fitted both
spectra simultaneously, without background subtraction, using \xspec. 
Our model  consisted of 
a single \apec\ plasma (to take account of the diffuse emission from
the galaxy; the ``source''),
plus background components. These comprised 
a power law with $\Gamma=1.41$ (to account for the hard X-ray background),
two \apec\ models with solar abundances and kT$=$ 0.2 and
0.07~keV (to account for the soft X-ray background) and, to model
the instrumental contribution, a broken power law
model and two Gaussian lines with energies 1.7 and 2.1~keV and negligible
intrinsic widths. We have found that this model
can be used to parameterize adequately the template background spectra. 

In order to disentangle the source and background components, 
given the general lack of
photons in these spectra, we tied the abundances and 
temperatures of the ``source'' \apec\ components between both
extraction regions, but allowed the normalizations to be free.
In the fainter galaxies the normalization of the source component in the 
``source free'' region tended, as expected, to zero. 
We also assumed that the background model 
normalization scales exactly
with the extraction area. In general, we found that this was
able to fit both spectra very well. In our subsequent spectral analysis, 
we did 
not background-subtract the data using the standard templates, but took 
into account the background by using an appropriately scaled version
of this model.

Even if the background spectrum varies substantially over the 
field-of-view, our background modelling is
most correct at largest distances from the source centroid, where the results
are most sensitive to the background. In fact, we found that 
the standard background templates fared much worse than these modelled
background estimates when  the data were from regions of low surface 
brightness.
We discuss this issue further, and how the choice of background
can affect our results in \S~\ref{sect_systematics_bkg}.

\subsection{\lx\ estimation} \label{sect_lx}
In order to provide a self-consistent analysis of the galaxies
in this sample, we obtained estimates of \lx\ based on the \chandra\
data. First, we estimated the flux 
within our chosen spectral extraction regions
from our the best-fitting spectral models
(\S~\ref{sect_spectroscopy}--\ref{sect_single_spectroscopy}).
Fluxes were computed separately for the gas and 
undetected point-sources in the energy-band 0.1--10.0~keV.
Since the adopted aperture will not contain all of the diffuse flux from the 
galaxy, we 
extrapolated the emission out to a projected radius of 300~kpc,
\ie\ the Virial radius for a $1.6\times 10^{12}$\msun galaxy.
We assumed spherical symmetry and parameterized the
surface brightness with a single or double $\beta$-model.
The $\beta$-model parameters were determined from fits to the 
radial surface brightness in the 0.3--2.0~keV band, using dedicated
software which can fold in the instrumental point-spread function,
which we computed at 1~keV.
Data from the vicinity of any detected point-sources were excluded from
the fit, and we assumed that the hot gas and undetected sources had
the same radial brightness distribution. 
The results of the surface brightness fits to each galaxy
will be discussed in detail in a subsequent paper. This procedure typically
corrects the flux upwards by a factor $\sim$1.1--4, depending on the 
shape of the surface brightness profile. Since it is by no means 
certain that this extrapolation is valid out to $\sim$300~kpc, we 
expect this to introduce some uncertainty into the estimated \lx. 
However, for our present purposes we believe this  approach is
sufficiently accurate.

To compute the total \lx\ of the galaxy, we also included the 
flux of all detected point-sources within the B-band
twenty-fifth magnitude (\dtwentyfive) isophote, all of which were
assumed to be associated with the galaxy for these purposes.
We fitted the composite spectrum of all these sources with our
canonical (kT=7.3~keV bremsstrahlung) model.
In most cases this gave a good fit to the data, although in a few
instances a single power law or power law plus disk blackbody
components were used instead to obtain a good fit. 
In the event a significant
low-luminosity AGN appears to be present in the galaxy 
(\ie\ NGC\thin 1553, IC\thin 4296), we omitted the flux from the AGN.
The total luminosity of the (detected, plus undetected) 
point-sources within \dtwentyfive\ was typically
in agreement with the estimate of \citet{kim03a}, based on
extrapolating the resolved X-ray luminosity functions in nearby early-type
galaxies. We found a mean 
\lx(point sources)/\lb$\sim 0.9\times 10^{30}$\ergps\lsun$^{-1}$,
in excellent agreement with these authors. The point-source populations
will be discussed in detail in a subsequent paper.

Comparing with the fluxes given in \citet{osullivan01a} we find
broad agreement, although our estimates tend to be
$\sim$0.25~dex higher. We attribute this discrepancy to differences
in spectral modelling and our 
extrapolation procedure.

%\clearpage

\section{Spectral analysis} \label{sect_spectra}
\begin{deluxetable*}{lrrrrrrrr}
\tabletypesize{\scriptsize}
\tablecaption{Emission-weighted average abundances\label{table_abundances}}
\tablehead{
\colhead{Galaxy} & \colhead{$\chi^2$/dof} & \colhead{\zfe} & \colhead{\zo/\zfe} & \colhead{\zne/\zfe} & 
\colhead{\zmg/\zfe} & \colhead{\zsi/\zfe} & \colhead{\zs/\zfe} & \colhead{\zni/\zfe} 
}
\startdata
\multicolumn{8}{c}{High-\lx\ galaxies}\\ \hline
% All backgrounds are modelled....... Best-fits are specified if possible..
% Note where I give two (almost) identical output lines for a given
% dataset, the one which is commented out is usually auto-generated
% from the syserr run...
IC\thin 4296& 54.3/68 & 1.8($>0.94$)& 0.0($<0.53$)& \ldots & 1.04$\pm0.47$ & 1.3$\pm0.6$ & \ldots &4.8($>0.2$) \\
 &  &[$^{+0.2}_{-0.1}$] &[+0.05] &\ldots &[$\pm 0.2$] &[$^{+0.03}_{-0.1}$] &\ldots &[$\pm1.7$] \\
% N507: 3D, 1T
NGC\thin 507 & 857/789 & 0.53$\pm0.07$$^\dagger$ & 0.0($<0.26$) & 0.52$^{+0.98}_{-0.52}$ & 0.81 $\pm$0.22 & $0.96\pm0.13$ & 1.51$\pm$0.30 & 2.77$^{+0.83}_{-0.70}$\\ 
 & &[$^{+0.14}_{-0.08}$] &[+0.6] &[$^{+0.26}_{-0.56}$] &[$^{+0.13}_{-0.50}$] &[$^{+0.1}_{-0.2}$] &[$^{+0.2}_{-0.4}$] &[$^{+0.7}_{-1.5}$] \\
% N 741 3D, 2T
NGC\thin 741 &$189.9/192$ & 1.19$^{+0.95}_{-0.62}$$^\dagger$ &$0.37^{+0.25}_{-0.30}$ &\ldots &$0.81^{+0.29}_{-0.48}$ &$0.98\pm 0.35$ &\ldots &$4.0^{+1.0}_{-2.0}$ \\
& &[$\pm 0.5$] &[$\pm 0.20$] &\ldots &[$\pm 0.15$] &[$^{+0.12}_{-0.05}$] &\ldots &[$^{+1.3}_{-0.8}$] \\
NGC\thin 1132 & 239.8/210 & 0.82$^{+0.29}_{-0.20}$ & 0.44$\pm0.37$ & \ldots & 0.70$\pm0.42$ & 1.26$^{+0.39}_{-0.34}$ & \ldots & 3.6$\pm1.4$\\
& &[$^{+0.12}_{-0.15}$] &[$^{+0.27}_{-0.18}$] &\ldots &[$^{+0.17}_{-0.29}$] &[$^{+0.08}_{-0.3}$] &\ldots &[$^{+1.2}_{-2.2}$] \\
% 1399 2d, 2t
NGC\thin 1399 & 1734/1223 & $1.19\pm0.10$$^\dagger$& $0.41\pm0.06$& $0.64\pm0.25$ & $0.76\pm0.08$ & $0.84\pm0.06$ & 1.14$\pm$0.12 & 2.61$\pm0.32$\\
% NGC 1600, 3D 2T
& &[$^{+0.6}_{-0.4}$] & [$^{+0.13}_{-0.11}$] & [$^{+0.05}_{-0.58}$] & [$^{+0.08}_{-0.22}$] & [$^{+0.05}_{-0.17}$] & [$^{+0.10}_{-0.49}$] & [$^{+0.4}_{-1.0}$]\\ 
NGC\thin 1600 & 151/158& 2.1$\pm1.3$$^\dagger$ & 0.10($<0.81$)& \ldots & 0.92$^{+0.77}_{-0.53}$& 0.85$^{+0.68}_{-0.36}$ & \ldots & 1.9($>0.3$)\\ 
&  &[$^{+1.3}_{-0.8}$] &[$^{+0.38}_{-0.10}$] &\ldots &[$^{+0.28}_{-0.13}$] &[$^{+0.38}_{-0.05}$] &\ldots &[$^{+1.2}_{-0.9}$] \\
% N4472: 3D, 1T
NGC\thin 4472&  785/740 & $1.4^{+1.7}_{-0.4}$$^\dagger$& 0.51$\pm$0.12 & 0.95 $\pm0.44$ & 1.02$\pm$0.11 & 1.25$\pm$0.11 & 2.36$\pm$0.33 & 3.28$\pm0.61$\\
& &[$^{+1.7}_{-0.6}$] &[$^{+0.07}_{-0.03}$] &[$^{+0.17}_{-0.28}$] &[$^{+0.07}_{-0.1}$] &[$^{+0.1}_{-0.2}$] &[$^{+0.3}_{-0.7}$] &[$^{+0.7}_{-1.8}$] \\
% This model is N5846 3D 2T
NGC\thin 5846  & 558/432 & $3.0^{+1.8}_{-0.5}$& 0.20$\pm0.13$ & 0.88$^{+0.61}_{-0.58}$ & 0.75$\pm 0.12$ & 0.78$^{+0.14}_{-0.11}$ & 1.14$^{+0.46}_{-0.38}$& 1.92$^{+1.81}_{-1.14}$\\
& &[$^{+0.3}_{-1.0}$] &[$^{+0.03}_{-0.04}$] &[$^{+0.17}_{-0.32}$] &[$^{+0.11}_{-0.05}$] &[$^{+0.17}_{-0.010}$] &[$^{+0.4}_{-0.2}$] &[$^{+2.5}_{-1.1}$] \\
%%% THIS 7619 is for the ACIS-I data....
%NGC\thin 7619 & 68.2/78 & 0.48$^{+0.30}_{-0.12}$ & 0.0($<0.76$) & \ldots & 0.84$\pm 0.41$ & 0.80$\pm0.30$ & \ldots & \ldots \\
%& &[$^{+0.52}_{-0.04}$] &[+2.1] &\ldots &[$\pm 0.21$] &[$^{+0.06}_{-0.25}$] &\ldots &\ldots \\
%%% THIS 7619 is for the ACIS-S data....
NGC\thin 7619 & 279/299 &  2.0$^{+2.5}_{-0.9}$   & 0.23$^{+0.37}_{-0.22}$ & \ldots & 1.06$\pm0.22$ & 0.91$\pm0.25$ & \ldots & 0.90$^{+1.43}_{-0.90}$ \\
 & & [$^{+3.0}_{-0.3}$] &[$^{+0.23}_{-0.04}$] &\ldots &[$^{+0.3}_{-0.1}$] &[$^{+0.10}_{-0.08}$] &\ldots &[$^{+1.10}_{-0.33}$] \\
NGC\thin 7626 & 59.6/45 & 0.37$^{+0.60}_{-0.16}$ & \ldots       & \ldots & 1.2$^{+0.88}_{-0.94}$ & 1.3$\pm1.0$ & \ldots & \ldots \\
 & &[$^{+1.30}_{-0.19}$] &\ldots &\ldots &[$^{+0.5}_{-0.4}$] &[$^{+0.9}_{-0.2}$] &\ldots &\ldots \\
\hline \multicolumn{8}{c}{Moderate-\lx\ galaxies}\\ \hline
NGC\thin 720$^1$  & 383.4/357& $0.80^{+0.45}_{-0.24}$ & 0.30$\pm 0.28$ & 0.68$\pm0.67$ & 1.26$\pm0.35$ & \ldots& \ldots & \ldots \\
              &          & [$^{+2.00}_{-0.35}$] &  [$^{+0.03}_{-0.30}$] & [$^{+0.06}_{-0.68}$] & [$^{+0.35}_{-0.16}$] & \ldots & \ldots & \ldots  \\
NGC\thin 1332$^1$ & 189.4/174& $1.2^{+1.9}_{-0.4}$ & 0.08$^{+0.13}_{-0.08}$ & 1.53$\pm0.31$ &1.07$\pm0.27$ & 0.87$\pm0.49$ & \ldots & \ldots \\
              &          & [$\pm0.3$]          & [$\pm0.4$]            & [$\pm0.1$]    &  [$\pm0.3$] & [$\pm0.2$] & \ldots & \ldots \\
NGC\thin 1387& 61.4/61& 0.38$^{+0.56}_{-0.15}$ & 0.32($<1.5$) & \ldots & 0.18$^{+0.46}_{-0.18}$ & 0($<$0.51)& \ldots & \ldots\\
&  &[$^{+1.4}_{-0.14}$] &[$^{+0.89}_{-0.32}$] &\ldots &[$^{+0.23}_{-0.04}$] &[+0.1] &\ldots &\ldots \\
NGC\thin 1407 & 222/221&  2.1$^{+1.1}_{-0.9}$$^\dagger$& 0.37$^{+0.21}_{-0.25}$ & \ldots & 1.10$\pm 0.23$ & 1.21$^{+0.31}_{-0.27}$ & 2.2$\pm1.1$ & 3.3$^{+1.7}_{-1.3}$\\
&  &[$^{+3.2}_{-0.3}$] &[$\pm 0.11$] &\ldots &[$^{+0.2}_{-0.09}$] &[$^{+0.2}_{-0.1}$] &[$^{+0.5}_{-1.2}$] &[$^{+0.7}_{-1.7}$] \\
NGC\thin 1549& 9.6/18 & 0.17($>0.06$) &   \ldots & \ldots & \ldots & \ldots & \ldots & \ldots \\
& &[$^{+4.0}_{-0.07}$] &\ldots &\ldots &\ldots &\ldots &\ldots &\ldots \\
% 1553, modelled background...
NGC\thin 1553& 35.8/35 & $0.14^{+0.09}_{-0.04}$ & 0.65$^{+0.47}_{-0.31}$ & 1.70$^{+0.55}_{-0.47}$ & \ldots &  \ldots & \ldots & \ldots \\
& &[$^{+0.14}_{-0.02}$] &[$^{+0.21}_{-0.16}$] &[$^{+0.1}_{-0.4}$] &\ldots &\ldots &\ldots &\ldots \\
% N1700 modelled. 1T
NGC\thin 1700& 33/36 & 5($>0.67$)& 0.31$^{+0.49}_{-0.31}$ & 0.70$\pm0.37$ & 0.64$\pm0.50$ & \ldots & \ldots & \ldots \\
&  &[-4.0] &[$\pm 0.16$] &[$^{+0.09}_{-0.45}$] &[$^{+0.25}_{-0.15}$] &\ldots &\ldots &\ldots \\
NGC\thin 3607& 43.0/41& 0.32$^{+0.60}_{-0.14}$ &  \ldots & \ldots & 0.63$\pm$0.63 & \ldots & \ldots & \ldots \\
& &[$^{+1.2}_{-0.11}$] &\ldots &\ldots &[$^{+0.29}_{-0.33}$] &\ldots &\ldots &\ldots \\
% N3923: 1T, mdled...
NGC\thin 3923 & 31.0/35& 1.15($>0.27$) & 0.12$^{+0.39}_{-0.12}$ & 1.48$^{+0.49}_{-0.39}$ & 0.85$\pm0.45$ & 4.1($>2.1$) & \ldots & \ldots \\
& &[$^{+3.4}_{-0.09}$] &[$^{+0.04}_{-0.01}$] &[$^{+0.1}_{-1.2}$] &[$^{+0.08}_{-0.19}$] &[$^{+0.2}_{-1.9}$] &\ldots &\ldots \\
NGC\thin 4365& 43.3/45& 5.0($>0.76$) & 0.46$^{+1.76}_{-0.36}$ & \ldots&  0.87$\pm0.88$& \ldots & \ldots &\ldots \\
%NGC4365 &$43.30/45$ &$5.0^{+0.}_{-0.8}$ &$0.46^{+0.72}_{-0.62}$ &\ldots &$0.87\pm 0.58$ &\ldots &\ldots &\ldots \\
 & &[$-3.4$] &[$^{+1.64}_{-0.02}$] &\ldots &[$^{+0.58}_{-0.77}$] &\ldots &\ldots &\ldots \\
NGC\thin 4552 & 139/143 & $0.71^{+0.27}_{-0.10}$ & 0.17$^{+0.11}_{-0.06}$ & $0.62\pm0.22$ & 0.76$\pm0.11$ & 1.1$\pm0.4$ & \ldots & 2.7$\pm$2.2\\
 & &[$^{+0.26}_{-0.15}$] &[$^{+0.21}_{-0.03}$] &[$\pm 0.15$] &[$^{+0.31}_{-0.04}$] &[$^{+0.3}_{-0.1}$] &\ldots &[$\pm 1.4$] \\
NGC\thin 5018& 23.0/24& 0.30($>0.14$) & \ldots & \ldots & \ldots & \ldots & \ldots & \ldots \\
& &[$^{+0.41}_{-0.11}$] &\ldots &\ldots &\ldots &\ldots &\ldots &\ldots \\
\hline \multicolumn{8}{c}{Low-\lx\ galaxies}\\ \hline
% N3115: mdl, 1T
NGC\thin 3115& 38.2/36& 0.26($>0.08$) & \ldots & \ldots & \ldots & \ldots & \ldots & \ldots \\
&  &[$^{+3.0}_{-0.15}$] &\ldots &\ldots &\ldots &\ldots &\ldots &\ldots \\
% N3585. Modelled 1T
NGC\thin 3585& 33.6/27& 0.63($>0.12$) &  \ldots & \ldots & \ldots & \ldots & \ldots & \ldots \\
& &[$^{+4.3}_{-0.49}$] &\ldots &\ldots &\ldots &\ldots &\ldots &\ldots \\
NGC\thin 3608& 33.8/23& 1.1($>0.14$)  & \ldots & \ldots & \ldots & \ldots & \ldots & \ldots \\
%N3608 &$38.59/24.00$ &$1.9\pm 3.3$ &\ldots &\ldots &\ldots &\ldots &\ldots &\ldots \\
& &[$\pm1.0$] &\ldots &\ldots &\ldots &\ldots &\ldots &\ldots \\
NGC\thin 4494 & 11.0/14& 0.28($>0.02$)& \ldots & \ldots & \ldots & \ldots & \ldots & \ldots \\
& &[$^{+4.7}_{-0.21}$] &\ldots &\ldots &\ldots &\ldots &\ldots &\ldots \\
NGC\thin 4621& 6/9& 0.31($>0$) & \ldots & \ldots & \ldots & \ldots & \ldots & \ldots \\
& &[$^{+4.7}_{-0.2}$] &\ldots &\ldots &\ldots &\ldots &\ldots &\ldots \\
NGC\thin 5845& 15.2/14& 0.62($>0.13$) & \ldots & \ldots & \ldots & \ldots & \ldots & \ldots \\
& &[$^{+4.4}_{-0.5}$] &\ldots &\ldots &\ldots &\ldots &\ldots &\ldots 
\enddata
\tablecomments{The best-fitting globally-averaged (see \S\ref{sect_spectroscopy}) emission-weighted abundances and abundance ratios  for each galaxy,
shown along with the quality of fit. Statistical errors represent the 90\% confidence region.
Figures given in square brackets are an estimate of the 
sensitivity of each measurement to possible sources of systematic error 
(see \S~\ref{sect_systematics}). Since we did not assess them for all
galaxies in the sample, calibration uncertainties
(\S~\ref{sect_systematics_calibration}) are not included in this error-budget
for most of the sample.
 These should certainly {\em not} be added in quadrature with
the statistical errors. Where we were able to obtain an abundance gradient (\S~\ref{sect_spectroscopy}), we estimated an emission-weighted \zfe, extrapolated
over a large aperture (see text); those affected galaxies are marked ($^\dagger$).
$^1$---results taken from \citet{humphrey04b}, corrected to our adopted 
abundance standard. Where parameters could not be constrained, they were fixed
at the Solar value, and listed as (\ldots).}
\end{deluxetable*}

%\clearpage

\subsection{Solar abundances standard} \label{sect_solar_standard}
Throughout this paper, we adopt the solar photospheric 
abundances of \citet{asplund04a}, which deviate significantly for many
of the key species from the previous standard of \citet{grsa}. 
The incorporation of detailed 3D  line transfer modelling (and, in
some cases, treatment of non-LTE conditions) has 
tended to reconcile so-called ``meteoritic'' and photospheric abundances
(for non-volatile species), so that discrepancies remaining are
typically at \ltsim 0.1~dex, \ie\ approximately the same level as the 
statistical uncertainties. We therefore adopt these abundances as our 
standard. This does, however, introduce some complications when comparing
with results reported in the literature since most authors adopt either 
the older abundances standard of \citet{grsa} or the outdated 
abundances of \citet{anders89}.

For comparison with our work, \zfe, \zo/\zfe, \zne/\zfe,
\zmg/\zfe, \zsi/\zfe, \zs/\zfe\ and \zni/\zfe\ referenced to the standard
of \citet{grsa} should be scaled by 1.12, 1.32, 1.55, 1.00, 0.98,
1.38 and 0.93, respectively.
Likewise, these abundances obtained with the outdated standard of
\citet{anders89} should be scaled by  1.66, 1.12, 1.07, 0.68,
0.66, 0.71 and 0.63, respectively.

\subsection{Error-bar estimation} \label{sect_errorbars}
It is worth noting that the error-bars generated simply with
\xspec\ tend to underestimate substantially the true confidence region.
This seems to arise because the fitting algorithms employed often
spuriously identify fit convergence (regardless of how low one sets the 
``critical delta-fit statistic'' parameter). This problem can be
mitigated by {\em re-starting} the fit a number of times from the 
apparent minimum and ensuring that the fit statistic has truly minimized.
Unfortunately the error-bar estimation routines assume that a single
iteration of the fitting algorithm is sufficient to characterize the 
shape of the $\chi^2$-space, which is seldom true. Since the failure to
converge results in an inflated $\chi^2$ for any given value
of an interesting parameter, this produces error-bars which are
systematically too small, sometimes dramatically so. 

We experimented
with two strategies to overcome this problem, which we found to agree
well in general. In our preferred technique, we used a script to 
emulate the behaviour of the \xspec\ ``error'' command, but minimizing
$\chi^2$
at each step with multiple iterations of the fit command (until the $\chi^2$
converged within a tolerance of 0.001). This has the advantage of 
identifying (small) local minima. The alternative method,
extensively employed in our previous papers 
\citep[\eg][]{buote03b} involves performing Monte-Carlo simulations
in which spectra are simulated from the best-fit model. The best-fit
model is then {\em fitted} to each simulated dataset and the 
confidence region inferred from the distribution of the best-fitting
parameter values.
Provided the fit statistic is well-described 
by the $\chi^2$  distribution, both techniques are statistically equivalent.

%\clearpage

\begin{figure*}
\centering
\plotone{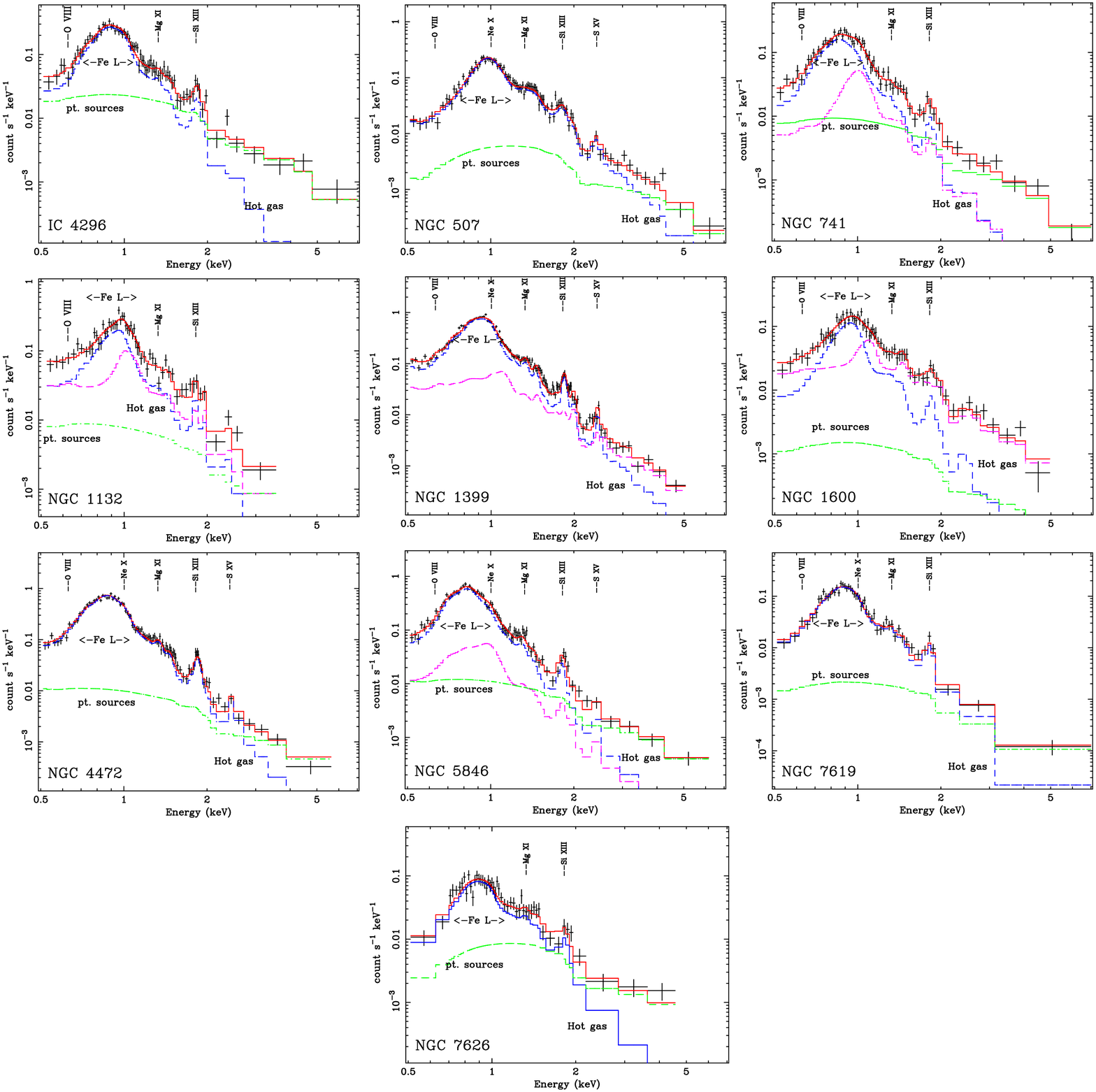}
\caption{Spectra of the high-\lx\ galaxies. Where spatially-resolved
spectroscopy was possible, only the central bin is shown, for clarity.
In addition we show the best-fit model folded through the 
instrumental response (solid line; red), and 
the contribution from undetected point sources (dash dot line; green)
and the one or two hot-gas components (dashed lines; blue and magenta).
The energies of interesting line features are included to 
guide the eye. \label{fig_spectra_hi}}
\end{figure*}

%\clearpage

\begin{figure*}
\centering
\plotone{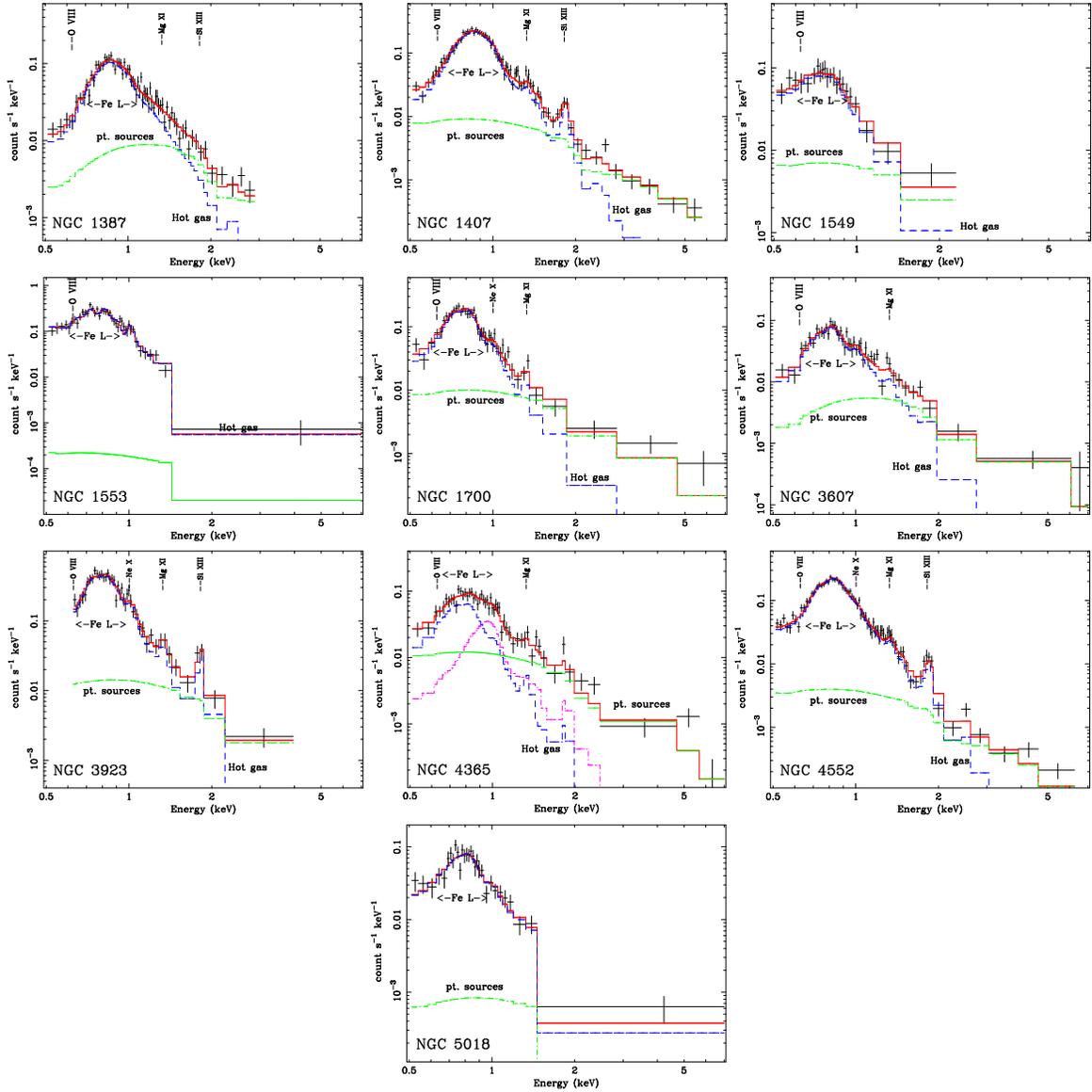}
\caption{Spectra of the moderate-\lx\ galaxies. Note that the point-source
component for NGC\thin 1553 and NGC\thin 5018 appears to be very small; 
however the  error-bars on the normalization of this component are 
considerable.
\label{fig_spectra_mid}}
\end{figure*}

%\clearpage

\begin{figure*}
\centering
\plotone{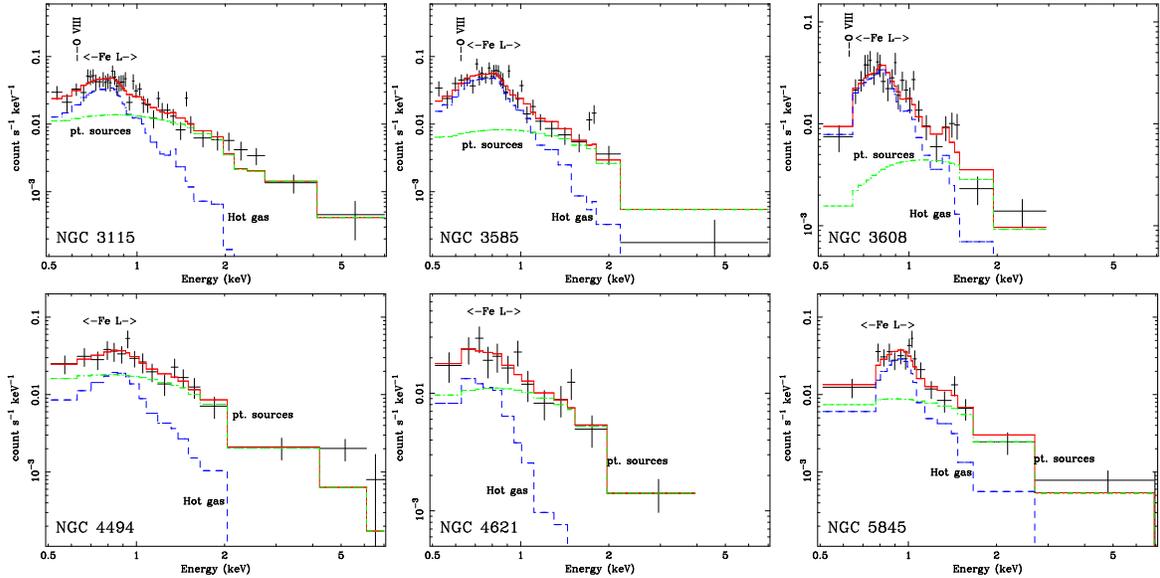}
\caption{Spectra of the low-\lx\ galaxies.\label{fig_spectra_lo}}
\end{figure*}

%\clearpage

\subsection{Spatially-resolved spectroscopy} \label{sect_spectroscopy}
\begin{figure*}
\plotone{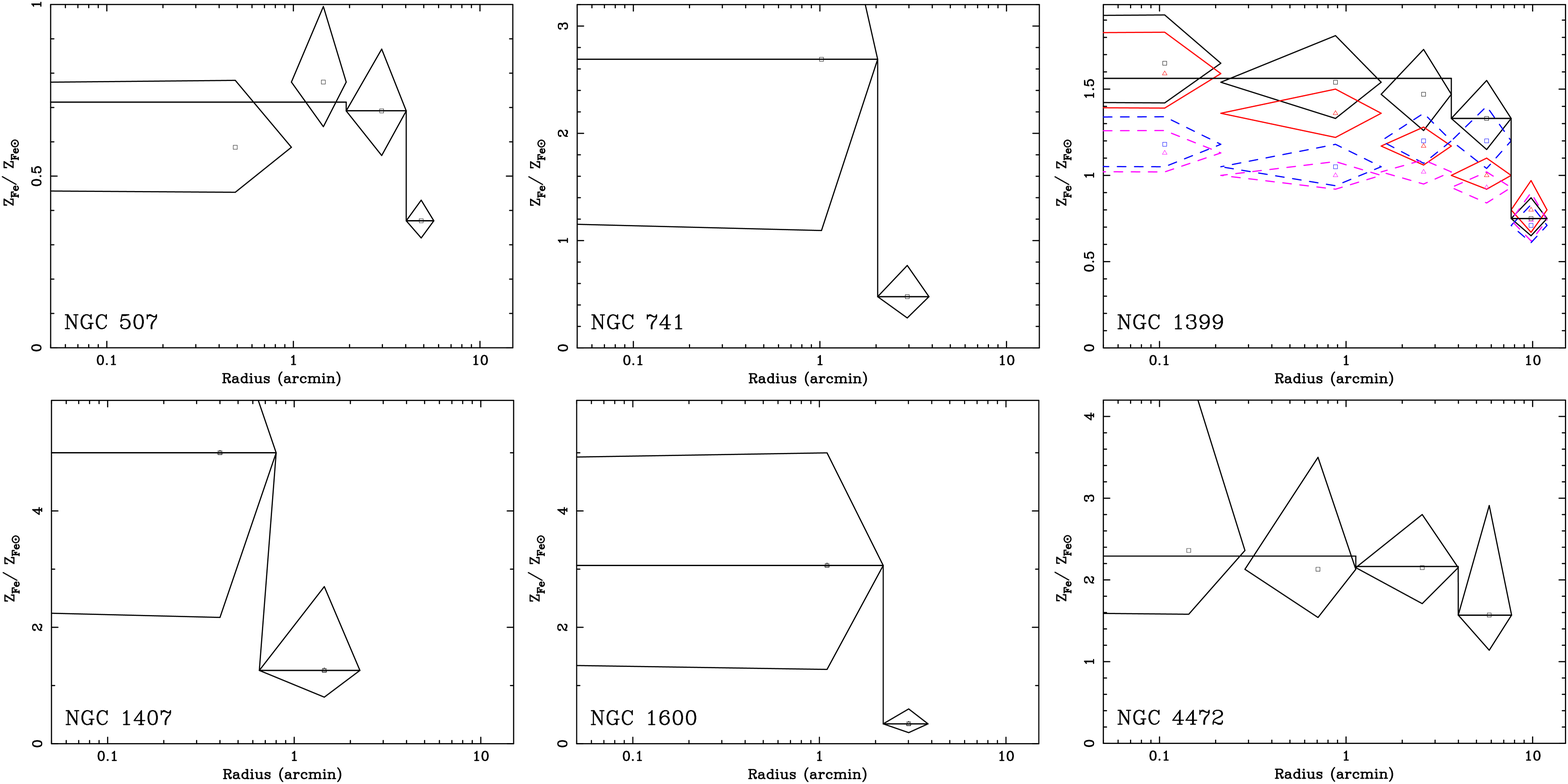}
\caption{The Fe abundance profiles, for those systems in which an abundance
gradient was measured. Also shown are the best-fitting models used to 
infer a global emission-weighted abundance
(\S~\ref{sect_spectroscopy}). To demonstrate the Fe bias and the 
importance of deprojection, we show separately for NGC\thin 1399
the profile obtained when 
two hot gas components were used in those annuli requiring them (solid lines)
and when only a single hot gas component is used (dotted lines). We also
show results for which deprojection was used (marked with squares) and for which
deprojection was not (marked with triangles). \label{fig_fe_profile}}
\end{figure*}

%\clearpage

Where the data were of sufficient quality, we extracted spectra in a number
of concentric annuli, centred on the nominal X-ray centroid. We determined
the centroid iteratively by placing a 0.5\arcmin\ radius aperture at the nominal
galaxy position (obtained from \ned) and computing the X-ray centroid
within it. The aperture was moved to the newly-computed centroid, and the 
procedure repeated until the computed position converged. Typically the 
X-ray centroid agreed with that from \ned. The widths of 
the annuli were chosen so as to contain approximately the same number of
background-subtracted photons and ensure there were sufficient photons in each
to perform useful spectral-fitting. We restricted the fits to the 
energy-band 0.5--7.0~keV, to minimize instrumental background, which
dominates at high energies, and to avoid calibration 
uncertainties at lower energies (however, see \S~\ref{sect_systematics_band}).
The spectra were rebinned to ensure a signal-to-noise ratio of at least 3
and at minimum 20 photons per bin (to validate $\chi^2$-fitting).

We fitted the
spectra using \xspec\ with 
a model comprising a hot gas ({\bf vapec}) component, plus an additional
7.3~keV bremsstrahlung component to take account of undetected 
point-source emission (this model gives a good fit to the detected
sources in nearby galaxies: \citealt{irwin03a}). We used a slightly 
modified form of the existing \xspec\
{\bf vapec} implementation so that \zfe\ is determined directly,
but for the remaining elements the fit parameters were the abundance 
{\em ratios} (in solar units) with respect to Fe. This model
enabled errors on the abundance ratios to be determined directly
from \xspec. In general, the data did not enable us to 
constrain any abundance {\em ratio} gradients, and so we tied the 
abundance ratios between all annuli.
The absorbing column density (\nh) was fixed at the Galactic value
\citep{dickey90}; the effect of varying \nh\ is discussed in 
\S~\ref{sect_systematics_nh}.
Where abundances could not be constrained, they were fixed at the 
Solar value.
We took account of possible hot gas spectral projection effects by employing
the {\bf projct} model implemented in \xspec, where possible. By fitting
the data in several annuli, it is possible to measure and, crucially,
constrain, any temperature gradients which contribute to the so-called
Fe bias.  We discuss the temperature profiles of each galaxy in detail in
a subsequent paper, and briefly in \S~\ref{sect_resolved_spectra}.
In order to improve the abundance constraints, which tend to be 
somewhat poorer than the temperature constraints, it was sometimes
necessary to tie the abundances between adjacent annuli (see below).
In the interests of physically meaningful results, we constrained
all abundances and abundance ratios to the range 0.0--5.0 times the
solar values.  Abundance ratios were kept fixed at the solar value
unless we could obtain interesting constraints during the fitting.
The spectra for the central annuli of all the galaxies are shown in 
Fig~\ref{fig_spectra_hi}--\ref{fig_spectra_mid}, along with the best-fitting
model. Table~\ref{table_abundances} lists globally-averaged
abundances derived from  each fit.

Since there is some evidence of limited multi-phase gas in some 
giant ellipticals and groups \citep{buote03a,buote02a,xue04a}, we experimented
with the addition of an additional hot gas component to the inner few
annuli of each galaxy. In a few 
cases (see below) this improved the fit significantly. 
It was generally not possible to determine
the abundances of this additional hot gas component separately,
and so they were tied to those of the other gas component in
the same annulus.
This two component model is a simple parameterization of 
multiple temperature gas components in the extraction aperture,
for example if there is a strong temperature gradient over the 
extraction region. It has been shown previously 
\citep[\eg][]{buote03b} that abundance constraints obtained
with this parameterization are generally in very good agreement
with those derived from more complex models which allow the emission-measure
to vary continously as a function of temperature.
In all cases discussed below
we included a component to account for undetected point-source emission.

To estimate an emission-weighted global abundance where there was 
some evidence of an abundance gradient, we first parameterised the 
radial surface brightness profile of each galaxy with a $\beta$-model 
fit. We then assumed that the underlying abundance profile is
well-described by a broken power law model. This model is a good fit to the 
profile seen in NGC\thin 1399, and is similar to that
in the centre of the group NGC\thin 5044
\citep{buote03b}. Where there was insufficient
data to constrain the radius of the break, we simply fixed it to
the midpoint of the innermost data-bin, inside which the 
abundance is flat (to prevent it inflating unphysically at small
radii). Using this model we were able to estimate the emission-weighted
abundance extrapolated to a ``global'' aperture
(for which we used 30\arcmin; the results
are relatively insensitive to this choice). 
For those galaxies with measured abundance profiles, we show the data and the 
best-fit model in Fig~\ref{fig_fe_profile}.

\subsubsection{Comments on individual galaxies} \label{sect_resolved_spectra}
{\em NGC\thin 507.}  We extracted spectra in seven contiguous,
concentric annuli, with outer radii 0.6, 1.0, 1.3, 1.9, 2.7, 4.0 and 
5.6\arcmin\ (14, 23, 31, 44, 63, 93 and 130~kpc), respectively. 
To improve the abundance constraints, we tied \zfe\ between the 
\first\ and \second\ annuli, the \third\ and \fourth\ annuli and the
\fifth\ and \sixth\ annuli.
The temperatures were allowed to vary in each annulus. The best-fitting
temperature profile rose from $\sim$1~keV in the centre to
$\sim$1.3~keV in the outermost radii. The abundance
profile is essentially flat, except for the outermost bin. The 
measured abundances generally agreed with  
\citet{kraft04a}, who used the \chandra\ data. 
We did not find any statistically significant
improvement in $\chi^2$ if a second hot gas component
was added ($\Delta \chi^2\sim$2). Nevertheless,
to effect a comparison with the \xmm\ results of \citet{kim04a},
we experimented with the addition of such a component to the 
innermost bin. 
We found that this increased the central \zfe\ to
$0.87^{+0.64}_{-0.28}$, which is marginally 
inconsistent with \zfe$\sim$2 found by these authors.
The reason for this discrepancy is unclear since
the temperatures of these components
(kT$\sim$0.8~keV and 1.4~keV) and the total flux of unresolved
point-sources within \dtwentyfive\ ($\sim 4\times 10^{-13}$\ergpscm),
which might systematically affect \zfe,
were very similar to those found by \citet{kim04a}.
Nonetheless, our abundance ratios were broadly consistent with
those reported for \xmm.

{\em NGC\thin 741.} We extracted spectra in three contiguous,
concentric annuli, with outer radii 0.6, 2.0 and 3.8\arcmin\
(12, 43 and 82~kpc), respectively. In order to obtain interesting
constraints, we tied \zfe\ between the inner two annuli. We found 
a statistically significant ($\Delta \chi^2$=10) improvement in the fit
if two, rather than one, hot gas components (kT$=$0.67~keV and 1.2~keV) 
were used in the  innermost bin; there is no evidence for the cooler 
component in the outer bins. 
The best-fitting abundances are somewhat higher than
\zfe=$0.37^{+0.37}_{-0.18}$ found by \citet{mulchaey03a}, who fitted
\rosat\ data with a single \mekal\ model (and no undetected source
component). We attribute the discrepancy to the Fe bias.

{\em NGC\thin 1132.} We extracted spectra in four contiguous, 
concentric annuli, with outer radii 0.8, 1.6, 2.4 and 3.7\arcmin\
(21, 43, 66 and 100~kpc), respectively. 
In order to obtain interesting constraints, we tied
the abundances in all four annuli. We found a $\sim$flat
temperature profile (kT$\simeq$1.0~keV) when only one hot gas 
component was used. We found a slight improvement
in the fit if two hot gas components (kT=0.8~keV and 1.6~keV)
were used in the central bin. Our best-fitting abundance within the 
4\arcmin\ radius was in good agreement with previous \asca\ 
determinations \citep{buote00a,mulchaey99a}. Recent \xmm\ measurements
show evidence of a significant abundance gradient 
\citep[][F. Gastaldello et al. 2005, in prep.]{gastaldello04a}; within 
$\sim$4\arcmin, our abundance determination is in good agreement with 
\xmm.

{\em NGC\thin 1399.} We extracted spectra in 10 contiguous, concentric
annuli, with outer radii
0.2, 0.7, 1.5, 2.4, 3.0, 3.6, 4.2, 5.4, 7.7 and 12\arcmin\ 
(1, 4, 8, 12, 16, 20, 22, 29, 41 and  64~kpc), respectively. 
In order to improve abundance 
constraints, we tied together \zfe\ between annuli 2 and 3, 
between annuli 4--6 and between annuli 7--9.
We found a significant
improvement in the fit if two hot gas components 
were used in the inner  6 bins. For the cooler component
kT rises from $\sim$0.7~keV to 1.3~keV. The hotter component
has kT$\sim$1.5~keV, but was less well-constrained.
Although the best-fitting
model was  not formally acceptable  
adding an additional hot gas component did not improve the fit
further. Given the excellent S/N of the data and the fact
that the fractional fit residuals are typically $\sim$ a few percent, 
it seems probable that the remaining errors are primarily systematic, for 
example calibration effects.
We found evidence of an abundance gradient, with a sharp drop-off
at $\sim$7\arcmin, as shown in Fig~\ref{fig_fe_profile}.
Our abundance profile is in excellent agreement with
that derived from \xmm\ data \citep{buote02a}. These results are
also in agreement with previous single-aperture \asca\ abundance 
measurements \citep[\eg][]{buote98c}.

{\em NGC\thin 1407.} We extracted spectra in 3 contiguous,
concentric annuli with outer radii of 0.3, 0.8 and 2.1\arcmin\
(2.7, 6.1 and 16~kpc), respectively. In order to improve
the constraints, the abundances were tied between the two inner
annuli. We only required a single hot gas component in each
annulus. The temperature rose from $\sim$0.65~keV in the centre
to 1.0~keV in the outer annuli. We find some evidence of an abundance gradient in the data,
although the error-bars are rather large.
Given the large errors on our interpolated global abundance, our
best-fitting values are in general agreement with the \asca\ value of 
\zfe$=0.88^{+1.63}_{-0.60}$ found by \citet{buote98c}. 

{\em NGC\thin 1600.} We extracted spectra in 3 contiguous, concentric
annuli with outer radii 0.8, 2.2 and 3.8\arcmin\ (13, 36 and 62~kpc),
respectively. In order to obtain interesting constraints, we tied
\zfe\ between the inner two annuli. We found a significant improvement in the 
fit ($\Delta \chi^2$=15) when two hot gas components (kT$=$0.86~keV 
and $\sim$3~keV), rather than one, were used in the central bin.
In the outer radii, only a single hot gas component (kT=1.5~keV)
was needed.
Using \chandra\ data \citet{sivakoff04a} reported 
\zfe$\sim$1.8 for a two-temperature fit 
within 1 effective radius, falling to $\sim$0.5 
at $\sim$3\arcmin\ (using our solar standard), which is consistent
with our results.

{\em NGC\thin 4472.} We extracted spectra in  9 contiguous,
concentric annuli with outer radii 0.3, 0.6, 1.1, 1.6, 2.2,
3.0, 4.0, 5.2 and 7.7\arcmin\ (1.3, 2.8, 5.0, 7.1, 9.6, 13, 17, 23, 
34~kpc), respectively. In order to improve the constraints, we 
tied together \zfe\ in annuli 2 and 3, and also in annuli 4--7, 
and annuli 8--9. Only a single hot gas component
was required by the data in any given radius. The temperature
profile rose from kT=0.67~keV in the centre to 1.3~keV in the outer
annuli. There is
some evidence of slight abundance gradient, although the
error-bars are rather large. Our measured
\zfe\ in each bin is somewhat larger than the values $\sim$0.5--1.1 
(converting to our solar standard) found
by \citet{finoguenov00a}. This is most likely a consequence 
of the Fe bias; there is a clear temperature gradient even within
the smallest annuli those authors attempt to use and
furthermore our results agree well with \citet{buote98c}.

{\em NGC\thin 4552.} We extracted spectra in 3 contiguous,
concentric annuli with outer radii 0.3, 1.2 and 2.4\arcmin\
(1.2, 5.2 and 10~kpc), respectively. We tied abundances 
between all annuli. Only one hot gas component
was required in any annulus to fit the data. There is some evidence
of a central temperature peak, similar to that found in 
NGC\thin 1332 \citep{humphrey04b}: we found kT=0.58~keV in the centre,
falling to $\simeq$0.33~keV by the next annulus.
Using \asca\ \citet{finoguenov00a} found
\zfe$<0.12$, when fitting the spectrum with a single hot gas
component. We attribute this discrepancy to the
Fe bias.
The abundance ratios reported by 
those authors also disagree with our measured values, which 
is presumably a related effect.

{\em NGC\thin 5846.} We extracted spectra in 6 contiguous,
concentric annuli with outer radii 0.4, 0.7, 1.7, 2.3, 3.9\arcmin\
(2, 4, 7, 10, 14 and 24~kpc), respectively. We found that fitting
a single hot gas component (plus undetected source component),
with kT rising from $\sim$0.6~keV in the centre to $\sim$1~keV in 
the outer radii, gave a rather  poor fit ($\chi^2$/dof=630/450).
Adding an additional hot gas component to the inner annuli
did not improve the fit. However, the fit was considerably
improved if such a component was added to the outermost 
annulus ($\Delta \chi^2$=60). By inspection, the image shows
considerable large scale structure, which complements the 
disturbed morphology on smaller scales
\citep{trinchieri02a}. To investigate this
further, we constructed a hardness map (using energy-bands 0.1--0.8~keV 
and 0.8--7.0~keV, separately smoothing each image by convolution
with a Gaussian of width 5\arcsec), which revealed that the gas temperature
within each annulus is not uniform.
We confirmed this explicitly for the outermost annulus by restricting
the extraction region to a narrow sector chosen to contain the 
softest photons; the best-fitting temperature of the gas in the
outer annulus was then significantly lower (falling to 0.8~keV). 
A detailed analysis of the two-dimensional temperature structure
of this galaxy is beyond the scope of this present work.
However it is sufficient to add  additional hot gas components in each annulus
for which the fit was significantly improved to obtain a 
reasonable estimate of \zfe.  The addition of 
a third hot gas component to each annulus did not improve the fit
significantly.
Although it is not clear that spherically-symmetric 
deprojection is appropriate in such a system, the fit was somewhat
better when {\em projct} was used than when it was omitted.
The final $\chi^2$ was poor; this may reflect systematic
errors in the response, since the source is bright, or it may
reflect the imperfect spectral deprojection resulting from
the spherical approximation, or it may be a consequence of the 
complicated temperature structure. 
We did not find any clear evidence of an abundance
gradient so the abundances of all gas components were tied. 

{\em NGC\thin 7619.} We extracted spectra in 7 contiguous, concentric annuli
with outer radii 0.2, 0.6, 1.1, 1.8, 2.5, 3.4 and 4.5\arcmin\
(2.5, 8.0, 16, 25, 35, 48 and 63 kpc), respectively. We did not find
strong evidence of a \zfe\ abundance gradient and so we tied abundances
between all annuli. Only one hot gas component was required in 
each annulus to fit the data. The temperature was seen to rise from 
$\sim$0.76 keV in the centre to $\sim$0.95 keV in the outer annulus.
Our measured \zfe\ is in excellent agreement with that reported
by \citet{buote98c}.

\subsection{Single aperture spectroscopy} \label{sect_single_spectroscopy}
In many cases there were insufficient photons to attempt spatially-resolved
spectroscopy. We therefore extracted the spectra from a single
aperture, the size of which was chosen crudely to maximize the 
S/N, and was typically $\sim$2--3\arcmin. We fitted our canonical
hot gas plus bremsstrahlung model to each spectrum. 
We note that, for low-\lx\ systems, this model is tantamount to the
two-temperature hot gas modelling used by \citet{buote98c}, who found that 
one component was typically hot (kT\gtsim 5~keV) in
the most X-ray faint systems. In some cases a
statistically significant improvement in the fit statistic 
($<1$\% chance the improvement is spurious, on the basis of an f-test)
was found if an additional hot gas component was included. 
We state below, for each galaxy, whether one or two hot gas components
 were required. All fits included an undetected point-source component.
The spectra for all the galaxies are shown in 
Fig~\ref{fig_spectra_hi}--\ref{fig_spectra_lo}, along with the best-fitting
model.
\subsubsection{Comments on individual galaxies}

{\em IC\thin 4296.} We extracted the spectrum from a 1.5\arcmin\ (22~kpc)
aperture, omitting data from the vicinity of the central LLAGN. 
We found that the addition of an extra hot gas component
significantly ($\Delta \chi^2$=11) improved the fit. The gas
components had kT=0.7~keV and 1.3~keV. Our results are
in excellent agreement with \zfe\ derived from \asca\
\citep{buote98c}. 

{\em NGC\thin 1387.} We extracted the spectrum from a 2\arcmin\ (11~kpc)
aperture. Only one hot gas component (kT=0.65~keV) was needed. Using \rosat\ 
\citet{jones97} measured \zfe=0.25$^{+0.47}_{-0.12}$, in agreement
with our results, even though these authors omitted a component to account
for point-source emission, which contributes $\sim$35\% of the flux in our
extraction aperture.

{\em NGC\thin 1549.} We extracted the spectrum from a 3\arcmin\ (16~kpc)
aperture. There was very little diffuse emission, but some evidence of metal
enrichment in the $\sim$0.4~keV gas. Using the \rosat\ PSPC, 
\citet{davis96b} reported 
\zfe$<0.18$, omitting a component to account for undetected point sources. 
Point sources contribute $\sim$50\% of the flux in our extraction region,
and so we attribute the discrepancy to incorrect modelling of these sources,
which could not be resolved with \rosat.

{\em NGC\thin 1553.} We extracted the spectrum from a 2\arcmin\ (10~kpc)
aperture,
excluding data in the vicinity of the central LLAGN. The \chandra\ image
reveals remarkable structure (a striking S-shaped pattern, which may
be related to interaction with the central AGN: \citealt{blanton01b}).
\citet{blanton01b} reported \zfe$=0.17^{+0.10}_{-0.08}$ for the hot gas,
using the \chandra\ data, which agrees very well with our fit with a single
$\sim$0.4~keV gas component. We do not find any evidence of a hard component,
such as reported by these authors, in addition to the hot gas and 
point source components. However, if we include a small amount of 
flaring-contaminated data, such a component is required.
Given the low abundance and the complexity of the ISM, we 
experimented with the addition of a second (kT$\sim$0.86~keV) gas component. 
It did not 
improve the fit appreciably ($\Delta \chi^2 \simeq2$), but \zfe\
was significantly increased to 0.24$^{+0.64}_{-0.11}$, and \zne/\zfe\
reduced to 0.67$^{+1.00}_{-0.57}$. The very low abundance
we obtain for this galaxy may, therefore, be mitigated by the Fe bias.
However the data do not require such an additional hot gas component.

{\em NGC\thin 1700.} We extracted the spectrum from a 1.5\arcmin\ 
(18~kpc) aperture. Only a single $\sim$0.4~keV gas component was required to 
fit
the data. Using \chandra\ data \citet{statler02a} reported
\zfe=0.83$\pm0.50$, omitting a component to account for undetected 
point-sources (which contribute$\sim$30\% of the flux in our extraction
region). Nonetheless, this value is in agreement with our results.

{\em NGC\thin 3115.} We extracted the spectrum from a 2\arcmin\
(5~kpc) aperture. The diffuse component was dominated by the undetected
point-source component. However, there is still some evidence of a 
single $\sim$0.4~keV gas component. The 
abundances were very poorly constrained.

{\em NGC\thin 3585 and NGC\thin 4494.} We extracted spectra for
both of these sources from 2\arcmin\ (11 and 9~kpc, respectively)
apertures. Both galaxies were studied with \xmm\ by 
\citet{osullivan04a}, who reported \zfe$<0.1$,
with tight abundance constraints. Such a low abundance is excluded
by our data for NGC\thin 3585, although it is consistent within
the very large error-bars we found for NGC\thin 4494. 
Inspection of the \chandra\ spectrum of NGC\thin 3585
clearly reveals the presence of the Fe hump, indicating
substantial Fe enrichment (Fig~\ref{fig_spectra_lo}). In contrast  
the hot gas component in NGC\thin 4494 is clearly overwhelmed by the 
point-source contribution. The lower spatial 
resolution of \xmm\ makes substantially more point-source contamination 
than for \chandra\ in  {\em both} spectra inevitable. This may explain
the discrepancy with our results, since the metal abundance of the hot gas 
will then be highly sensitive to the ability to 
model the spectrum of the undetected sources.
Although {\em on average} the composite spectra of detected
sources in \chandra\ fields can be well-approximated by a simple 
bremsstrahlung or power law model \citep{irwin03a}, there is no reason to
believe that this is the exact spectral model for the undetected 
sources seen in any given galaxy (see \S~\ref{sect_systematics_sources}).
This is problematical where the gas parameters depend sensitively 
on its shape. Since \zfe$<0.1$ is difficult to understand in terms of 
the current picture of metal enrichment, but is more consistent with
what we might expect if  undetected sources are not 
properly accounted for, it seems 
likely that our best-fitting values are more representative of the 
true abundances in these systems.

{\em NGC\thin 3607.} A spectrum  was extracted from a 2\arcmin\ (12~kpc)
aperture. There is significant diffuse gas, clearly showing the characteristic
Fe L-shell ``hump''. The data only required one kT=0.46~keV gas
component, and \zfe\ is moderately well-constrained. \citet{matsushita00a}
reported a similar \zfe\ using \asca\ data.

{\em NGC\thin 3608.} A spectrum was extracted from a 2\arcmin\ (12~kpc)
aperture. Only one $\sim$0.35~keV gas component was required to fit the 
data. The hot gas contributes $\sim$50\% of the flux in the extraction
aperture, but the abundances were poorly constrained. 

{\em NGC\thin 3923.} A spectrum was extracted from a
1.5\arcmin\ (9~kpc) aperture. We restricted spectral-fitting to the
0.6--4.0~keV band, since there was some evidence of features
outside this range which may be an artefact of the prolonged mild
flaring in this observation, which it was impossible entirely to excise. Only one 
$\sim$0.38~keV gas component was needed, in addition to undetected
sources. The best-fitting \zfe\ was poorly-constrained, but in
good agreement with previous \asca\ measurements \citep{buote98c}.
Intriguingly, the \zsi/\zfe\ ratio was significantly higher than in
the other systems, since producing the prominent Si line evident
in the spectrum  (Fig~\ref{fig_spectra_mid})
in such cool plasma requires an extremely high metallicity.
This is not a very high significance effect, since an
acceptable fit ($\chi^2$/dof=38/36) 
can still be obtained if we constrain \zsi/\zfe$=1$. 
An alternative explanation might be the presence of multiple temperature components
in the aperture, in which case our inferred temperature may not 
be representative. There was some indication
of an improved fit if an additional $\sim$0.85~keV gas component
was included, but the improvement was not significant
($\Delta \chi^2$=3). Fitting two temperatures did result in
a smaller Si/Fe ratio (2.2$^{+2.2}_{-1.1}$), which is large but marginally 
consistent with solar.

{\em NGC\thin 4365.} We extracted a spectrum from a 2\arcmin\ 
(11~kpc) aperture. We found two hot gas components (0.36 and 0.93~keV)
were required by the data, and the resulting \zfe\ was poorly-constrained.
\citet{sivakoff03} reported similarly poorly-constrained abundances
for a similar region, using \chandra\ data. 
These authors also report some evidence of an abundance gradient.
However, given the challenges of background subtraction in such
a low surface-brightness regime, and the few available photons,
we did not find compelling evidence.

{\em NGC\thin 4621.} We extracted a spectrum from a 2\arcmin\
(10~kpc) aperture. There is very little diffuse emission, and the 
undetected point source component dominates the spectrum, however
a $\sim$0.3~keV gas component was needed.
The abundances could not be constrained well.

{\em NGC\thin 5018.} We extracted a spectrum from a 2\arcmin\ (25~kpc)
aperture. Only one $\sim$0.5~keV gas component was required to fit the data.
The abundances were poorly constrained. 

{\em NGC\thin 5845.} We extracted a spectrum from a 2\arcmin\ (14~kpc)
aperture. There was very little diffuse emission, but some evidence of
Fe-enriched $\sim$0.83~keV gas. 

{NGC\thin 7626.} We extracted a spectrum from a
2\arcmin\ (29~kpc) aperture. 
Only a single hot gas component, with kT=0.74~keV, 
was required. 
Our best-fitting \zfe\ were in good agreement
with those determined using \asca\ \citep{buote98c}.

\section{Systematic errors} \label{sect_systematics}
In this section, we address the extent to which systematic uncertainties
may impact upon our results. 
An estimate of the uncertainty due to these effects 
is given for each galaxy in Table~\ref{table_abundances}. These numbers
reflect the sensitivity of the best-fitting parameter values to 
each source of potential error, and we stress that they should certainly
{\em not} be added in quadrature with the statistical errors. 
We discuss a number of effects in detail below. 
Those readers uninterested in the technical details of the analysis 
may like to proceed directly to \S~\ref{sect_discussion}.
A breakdown of the systematic error-budget for three representative
galaxies is shown in Table~\ref{table_syserr}. These are chosen approximately
to span the luminosity, temperature and metallicity range of those galaxies
in our sample for which interesting constraints were found.

%\clearpage

\begin{deluxetable*}{lllllllllll}
\tablecaption{Abundance measurement error budget\label{table_syserr}}
\tabletypesize{\scriptsize}
\tablehead{
\colhead{Par.} & \colhead{value} &
\colhead{${\rm \Delta}$Stat.} &
\colhead{${\rm \Delta}$calib} & \colhead{${\rm \Delta}$bkd}
& \colhead{$\Delta$code} & \colhead{$\Delta$projection} &
\colhead{$\Delta$bandw.}& \colhead{$\Delta$sources} &
\colhead{$\Delta$\nh} & \colhead{$\Delta$Fe bias}\\}
\startdata
\multicolumn{11}{l}{NGC\thin 1399} \\\hline
\zfe & 1.18 & $\pm0.08$ & -0.1 & $\pm0.2$ & +0.27 & $\pm$0.01 & $^{+0.6}_{-0.4}$ & $+0.3$ & -0.3 & -0.3 \\
\zo/\zfe & 0.41 & $\pm0.06$ & +0.02 &  $\pm0.02$ & -0.11 & $\pm0.01$ & +0.13 & $\pm 0.03$&  +0.03 & +0.03 \\
\zne/\zfe & 0.64 & $\pm0.25$ & -0.17 & $\pm0.10$  & -0.58 & -0.04 & $^{+0.04}_{-0.21}$ & $^{+0.05}_{-0.14}$ & -0.12 & -0.24 \\
\zmg/\zfe & 0.76 & $\pm0.08$ & +0.08 & $^{+0.08}_{-0.04}$ & -0.22 & +0.02 & -0.17 & $\pm0.02$ & -0.17 & +0.10 \\
\zsi/\zfe & 0.84 & $\pm0.06$ & +0.04 & +0.05 & -0.17 & +0.02 & -0.07 & $\pm0.02$ & -0.08 & +0.11 \\
\zs/\zfe  & 1.14 & $\pm0.12$ & -0.49 & -0.07 & -0.38 & $\pm0.01$ & $\pm0.1$ & $\pm0.01$ & -0.1 & +0.2 \\
\zni/\zfe & 2.61 & $\pm0.32$ & -0.86 & $^{+0.4}_{-0.1}$ &  -1.0  & +0.1 & -0.2 & $\pm$0.2 & -0.3 & +0.7\\ 
\hline \multicolumn{11}{l}{NGC\thin 4552} \\\hline
\zfe & 0.71 & $^{+0.26}_{-0.16}$ & -0.30 & $^{+0.02}_{-0.07}$ & $-0.04$ & $\pm0.02$ & $^{+0.35}_{-0.16}$ & -0.07& +0.05 & +0.20\\
%fe &$0.74$ &$0.74^{+0.31}_{-0.21}$ &\ldots &$0.74^{+0.01}_{-0.02}$ &$0.74^{+0.}_{-0.04}$ &$0.74$ &$0.74^{+0.35}_{-0.16}$ &$0.74^{+0.}_{-0.07}$ &$0.74^{+0.05}_{+0.}$  \\
\zo/\zfe  & 0.17 & $^{+0.11}_{-0.06}$ & +0.11 & $\pm0.03$ & +0.07 & $\pm0.01$ & +0.21 & +0.02 & $\pm$0.01 & -0.03\\
%o &$0.16$ &$0.16\pm 0.11$ &\ldots &$0.16\pm 0.02$ &$0.16^{+0.06}_{+0.}$ &$0.16$ &$0.16^{+0.20}_{+0.}$ &$0.16^{+0.02}_{+0.}$ &$0.16\pm 0.010$  \\
\zne/\zfe & 0.62 & $\pm0.22$ & +0.99 & $\pm0.02$ & +0.10& $\pm0.01$ & $^{+0.15}_{-0.06}$ & +0.06 & +0.04  & +0.04\\ 
%ne &$0.74$ &$0.74^{+0.21}_{-0.18}$ &\ldots &$0.74^{+0.02}_{-0.03}$ &$0.74^{+0.}_{-0.25}$ &$0.74$ &$0.74^{+0.14}_{-0.06}$ &$0.74^{+0.06}_{+0.}$ &$0.74^{+0.05}_{+0.}$ \\
\zmg/\zfe & 0.76 & $\pm0.10$ & +0.30 & $^{+0.03}_{-0.05}$ & +0.34 & $\pm0.01$ & $^{+0.13}_{-0.04}$ & +0.05 & +0.06 & +0.06\\
%mg &$0.78$ &$0.78\pm 0.12$ &\ldots &$0.78\pm 0.03$ &$0.78^{+0.31}_{+0.}$ &$0.78$ &$0.78^{+0.13}_{-0.03}$ &$0.78^{+0.05}_{+0.}$ &$0.78^{+0.06}_{+0.}$ \\
\zsi/\zfe & 1.1 & $\pm0.4$ & $^{+0.05}_{-0.02}$ & $\pm 0.01$ & +0.2 & $\pm0.01$ & $^{+0.3}_{-0.1}$ & +0.07 & +0.1  & +0.02\\
%si &$1.1$ &$1.1\pm 0.4$ &\ldots &$1.12^{+0.02}_{+0.}$ &$1.1^{+0.3}_{+0.}$ &$1.1$ &$1.1^{+0.3}_{-0.1}$ &$1.12^{+0.07}_{+0.}$ &$1.12^{+0.06}_{+0.}$  \\
\zni/\zfe & 2.7 & $\pm2.2$ & -0.95 & +0.3 & -1.4 & $\pm0.2$ & $\pm1.3$ & -0.04 & $\pm0.01$ & -0.7\\
%ni &$2.2$ &$2.2\pm 2.5$ &\ldots &$2.2^{+0.2}_{+0.}$ &$2.2^{+0.}_{-0.7}$ &$2.2$ &$2.2^{+0.8}_{-1.1}$ &$2.2^{+0.}_{-0.4}$ &$2.17\pm 0.010$  \\
\hline \multicolumn{11}{l}{NGC\thin 3607} \\\hline
\zfe &$0.32$ &$^{+0.60}_{-0.14}$ &+0.08 &$^{+0.04}_{-0.11}$ &-0.06 &\ldots &$^{+1.2}_{-0.11}$ &-0.06 &$\pm 0.010$ & +0.06 \\
\zmg/\zfe &$0.63$ &$\pm0.63$ &-0.33 &-0.29 &+0.28 &\ldots &$^{+0.04}_{-0.26}$ &+0.09 &+0.02 & +0.06 \\
\enddata
\tablecomments{Emission-weighted abundance measurements for three
representative galaxies chosen from the sample.
In addition to the 90\% statistical
(Stat.) errors, we present an estimate of the possible
magnitude of uncertainties on the abundances due to
calibration (calib), background (bkd), plasma code (code),
projection effects (projection), bandwidth (bandw.) and
undetected source spectrum
(sources). 
We also include an assessment of the impact of the ``Fe bias''
($\Delta$Fe bias). For NGC\thin 1399, this is the effect of fitting a 
only one gas component to each annulus (see text). For NGC\thin 4552 and NGC\thin 3607
it is the effect of adding an extra hot gas component to the central bin.
It is difficult to assess the impact of multiple
sources of systematic error {\em simultaneously} affecting our results
(and it certainly is not correct simply to combine the errors in
quadrature), so it is probably most correct to assume that the source
of the largest systematic error dominates the systematic uncertainties.
The magnitudes of the systematic errors listed are the changes in best-fit
values, which can be more sensitive to such errors
than the confidence regions.
}
\end{deluxetable*}

%\clearpage

\subsection{Calibration issues} \label{sect_systematics_calibration}
Uncertainties in the absolute calibration of \chandra\ may impact on our 
ability to
determine reliable abundances for some or all of the sample. For example,
it is not possible in standard processing to correct data from the 
ACIS-S3 chip for ``charge transfer inefficiency'' (CTI).
\citet{townsley02} proposed
an algorithm to compensate for this effect, which
we previously found to have a noticeable effect on \zfe\ of 
NGC\thin 1332
($\Delta$\zfe$\simeq$0.3) \citep{humphrey04b}.
Unfortunately it was not clear to what extent the change arose
due to the effects of CTI or the older calibration of that technique.

As an alternative means of assessing the calibration uncertainty of
\chandra, we instead reduced \xmm\ archival observations of 
several galaxies. Since
\xmm\ is calibrated independently from \chandra, any discrepancies
will be indicative of the magnitude
of possible calibration errors. 
In the interests of efficiency, we only considered three galaxies
for this analysis,
NGC\thin 1399, NGC\thin 4552 and NGC\thin 3607, which reasonably
span the temperature, metallicity and luminosity range
in our sample.  A fully self-consistent analysis combining \xmm\ and 
\chandra\ data is beyond the scope of this present work, and 
will be addressed in a future paper.
The \xmm\ data were processed following a standard procedure outlined
in F. Gastaldello et al.\ (2005, in prep.) and spectra were extracted from
similar apertures to those used in the \chandra\ abundance analysis, 
where possible.

For NGC\thin 1399, we extracted spectra from 5 concentric annuli,
with outer radii 1.5, 3.3, 5.5, 8.6 and 12\arcmin. We fitted the 
same best-fitting model as used for the \chandra\ analysis. We obtained
best-fitting abundances in very good agreement with the \chandra\ data,
with the exception of \zs/\zfe, which we found to be significantly
lower ($\sim$0.65).

For NGC\thin 4552, we used a single 1.25\arcmin\ aperture,
to which we fitted two hot gas components(since a temperature gradient
was evident in the \chandra\ data), plus an undetected source
model. The temperatures of the components were
in good agreement with the range of kT resolved by 
\chandra. We found a significant reduction in the
\zfe\ ($\Delta$\zfe$\simeq$0.3), comparable to that seen when CTI-correction
was applied to the NGC\thin 1332 data.
We also found significant increases in \zo/\zfe, \zne/\zfe\
and \zmg/\zfe, although these may be tied in to the reduction in \zfe. 

For NGC\thin 3607, we extracted data from a single 1.7\arcmin\ aperture
(avoiding a chip-gap). We found good agreement with our measured \chandra\
abundances, although the \xmm\ constraints were somewhat poorer.
The \xmm\ gas temperature was slightly higher (kT$\sim$0.55~keV), although
we found that the temperature of this system determined with \chandra\ was sensitive to 
the treatment of the background (more so than \zfe), 
which probably explains this discrepancy.

\subsection{Background modelling} \label{sect_systematics_bkg}
One of the chief sources of systematic uncertainty in
fitting extended, low surface-brightness objects is the treatment 
of the background. 
In our default analysis, we modelled the background, in a procedure
akin to that adopted by \citet{buote04c}. We consider this to be 
the most robust approach, since it implicitly
takes into account long-term variations 
in the non X-ray background (which are  typically $\sim$10\%, in 
the absence of flaring\footnote{\href{http://cxc.harvard.edu/contrib/maxim/acisbg/COOKBOOK}{http://cxc.harvard.edu/contrib/maxim/acisbg/COOKBOOK}}),
and field-to-field variations in the X-ray background. Furthermore,
it provides a natural means of subtracting off any extended, unrelated
emission, such as cluster emission for galaxies in a cluster environment.

To assess the sensitivity of our results to our 
background treatment, we experimented with altering the 
background normalization
by $\pm$15\%, comparable to the variation between
background fields. We found little impact on our results
for the highest surface-brightness regions, as might be expected.
In contrast, however, the results for the faintest systems were rather
sensitive to the background level. 

We also experimented with using the standard
\chandra\ background templates.
We found that they typically were not sufficiently accurate to 
be of use in very low surface-brightness regimes.
For example, for the outer two annuli of NGC\thin 1399,
we found that using the templates led to a substantial
over-subtraction of the background, giving a
poorer fit ($\Delta\chi^2$=128) and much higher 
abundances in these annuli ($\Delta$\zfe$\sim$0.3--0.5),
in disagreement with the \xmm\ abundance profile \citep{buote02a}.
We reiterate that the common practice of renormalizing the background
templates to match the high-energy (\ie\ non-X ray background)
is potentially dangerous since it will simultaneously
renormalize the (uncorrelated) cosmic X-ray background and 
instrumental lines.

\subsection{The plasma codes} \label{sect_systematics_mekal}
In order to estimate the extent to which the Fe L-shell modelling
 may impact upon our results, we experimented with replacing
the (default) \apec\ plasma code with a \mekal\ model. As has previously been 
seen \citep[\eg][]{buote03b}, we found that this change
led to slightly lower best-fitting \zfe\ where only a single temperature
hot gas component was required in any annulus, and a slightly higher 
\zfe\ where two temperatures were required. We found a complementary
effect upon the abundance ratios, which may reflect in part the 
changing \zfe. These changes do not  affect significantly the 
interpretation of our results. Furthermore, we typically found
that the \mekal\ model gave slightly poorer  fits to the data,
justifying our preference for \apec. 

\subsection{Bandwidth} \label{sect_systematics_band}
Presumably in part due to slight systematic errors in the response matrices,
it is well-established that the choice of bandwidth affects the best-fitting
abundances \citep[\eg][]{buote00c}. We investigated the magnitude
of this effect in our data by altering the bandwidth and re-fitting
the data. We adopted three different pass-bands
in addition to the nominal 0.5--7.0~keV band, specifically 0.7--7.0~keV,
0.5--2.0~keV and 0.4--7.0~keV. Although the response below $\sim$0.5~keV
is suspect, since most of the galaxies in our sample were observed
relatively early in the life of \chandra\ we expect that the degradation of 
the low-energy response, which is in part responsible for this uncertainty,
should not be dramatic. 
Since the strongest O lines are in the range 0.6--0.8~keV and the S line
is at $\sim$2.4~keV, we did not attempt to assess the impact of
adopting the 0.7--7.0~keV or 0.5--2.0~keV bands, respectively, on the
abundances of these species.

First considering the 0.5--2.0~keV band, we found  there are 
insufficient counts in the vicinity of the Fe L-shell
clearly to disentangle the hot gas and point-source components in the 
faintest systems. In this case \zfe\ was
typically very poorly-constrained. In the brighter systems, the Fe L-shell
is well-defined, which enables both temperature and metallicity to
be determined reasonably, although degeneracy with the point-source
component still persists, resulting in compromised error-bars
and higher \zfe. 
Given the poor constraints we did not include the magnitude of this 
effect in our systematic error tables.

For the 0.4--7.0~keV band, we tended to find a slight increase in 
the measured \zfe, whereas in contrast, for the 0.7--7.0~keV band,
we found a  significant reduction. Both of these results can
be understood in terms of the importance of constraining the continuum
at energies below the Fe L-shell ``hump''. If the continuum is not 
properly estimated, the equivalent widths of the lines will be in
error, resulting in an incorrect abundance determination. Since the 
errors in the continuum measurement will impact all equivalent widths
in the same sense, the abundance ratios are less affected by this
uncertainty.

\subsection{Undetected sources} \label{sect_systematics_sources}
We included a component to account for undetected
point-sources in our sample. The adopted model has been
shown to be a good approximation to the composite spectrum of the 
{\em detected} point-sources in a range of early-type galaxies
\citep{irwin03a}. It is by no means certain, however, that this 
model will provide a perfect description of the {\em undetected}
X-ray point-sources in any galaxy. This especially may be the case
since the spectra of fainter LMXB (which would be undetected)
tend to be harder than those of brighter objects \citep{church01}. This
difference may be exacerbated if a substantial part of the 
bright X-ray binaries are high/ soft state black hole systems. 

To estimate the sensitivity of our results to this effect, we 
adopted two tests. Firstly  we allowed the temperature of the 
bremsstrahlung component
to be fitted freely. In the brightest systems, this best-fitting
temperature tended to be slightly lower than the canonical value
($\sim$2--3~keV). This may arise due to a slight inadequacy in our
modelling of the hot gas; it tends to produce slightly higher
best-fitting abundances, {\em via} a similar mechanism to 
the Fe bias. The abundance ratios were not significantly 
altered by this test. In the faintest systems, the temperature 
tended, if anything, to increase, which produces a slight systematic
reduction in the best-fitting \zfe. 
As a second test, we simply omitted the point-source
component entirely. In the brightest systems, where the hot 
gas overwhelms the point-source contribution, the impact is 
negligible. In the fainter systems the omission of the point
source component leads to a dramatically poorer fit and 
substantially lower abundances. This is  easily understood
as the continuum level is artificially raised to fit the 
high-energy residuals, systematically reducing the line equivalent
widths.

\subsection{Hydrogen column-density} \label{sect_systematics_nh}
Another source of systematic uncertainty in our analysis 
is an error  in our adopted  hydrogen column to the galaxy.
We experimented with allowing the \nh\ to fit freely for each
galaxy. We found a strong anti-correlation between \zfe\ and \nh,
since the   abundance measurement 
is sensitive to the continuum  at low
energies. If \nh\ is under-estimated, the continuum level
below the Fe L-shell will also be under-estimated, artificially
raising \zfe. We found that the abundance ratios were relatively
unaffected by this uncertainty.

Typically the best-fitting \nh\ was in good agreement with our
adopted Galactic value. 
In a few cases the measured value was slightly higher,
although we do not believe this indicates substantial intrinsic cold
absorption in these galaxies, and is probably more representative of slight
systematic uncertainties in the modelling and the responses
 at the lowest energies. 

\subsection{Fe bias} \label{sect_systematics_fe_bias}
Although we have endeavoured to remove the Fe bias wherever possible, there
exists the possibility that the quality of the data in some cases prevents
us from detecting the presence of multi-temperature hot gas. We therefore
include in our error-budget calculation a component to account for the Fe
bias, if appropriate. In Table~\ref{table_syserr}, for those galaxies
which require only one hot gas component, we show the impact of adopting 
a two-temperature model. For those galaxies requiring a two-temperature
model, we show the impact of fitting a single hot gas component model.
In order to make the systematic error-bars as meaningful as 
possible in the total error-budget shown in Table~\ref{table_abundances}, 
however, we only include the Fe bias term for systems requiring just one 
 hot gas model.

\section{Discussion} \label{sect_discussion}
\begin{figure*}
\centering
\plottwo{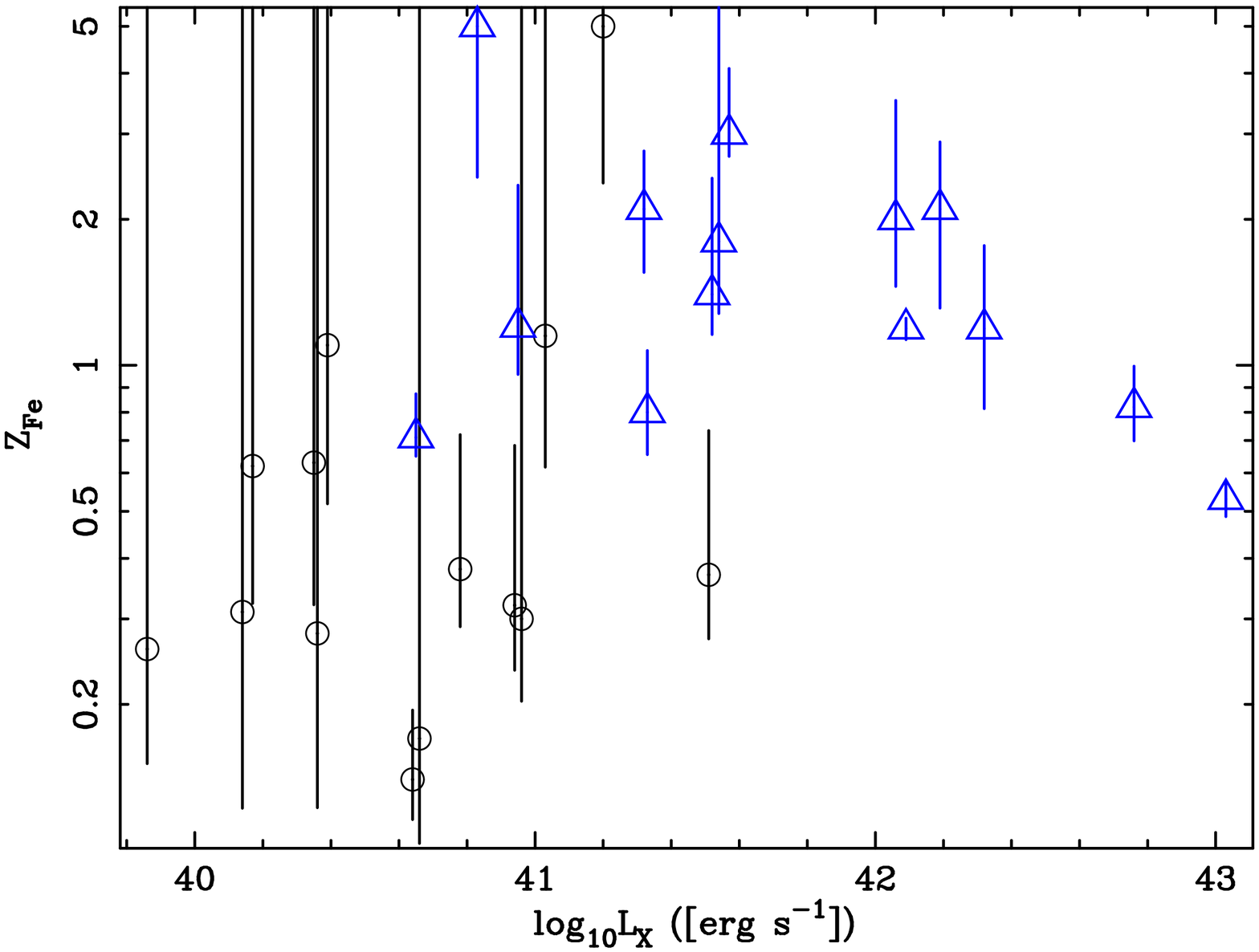}{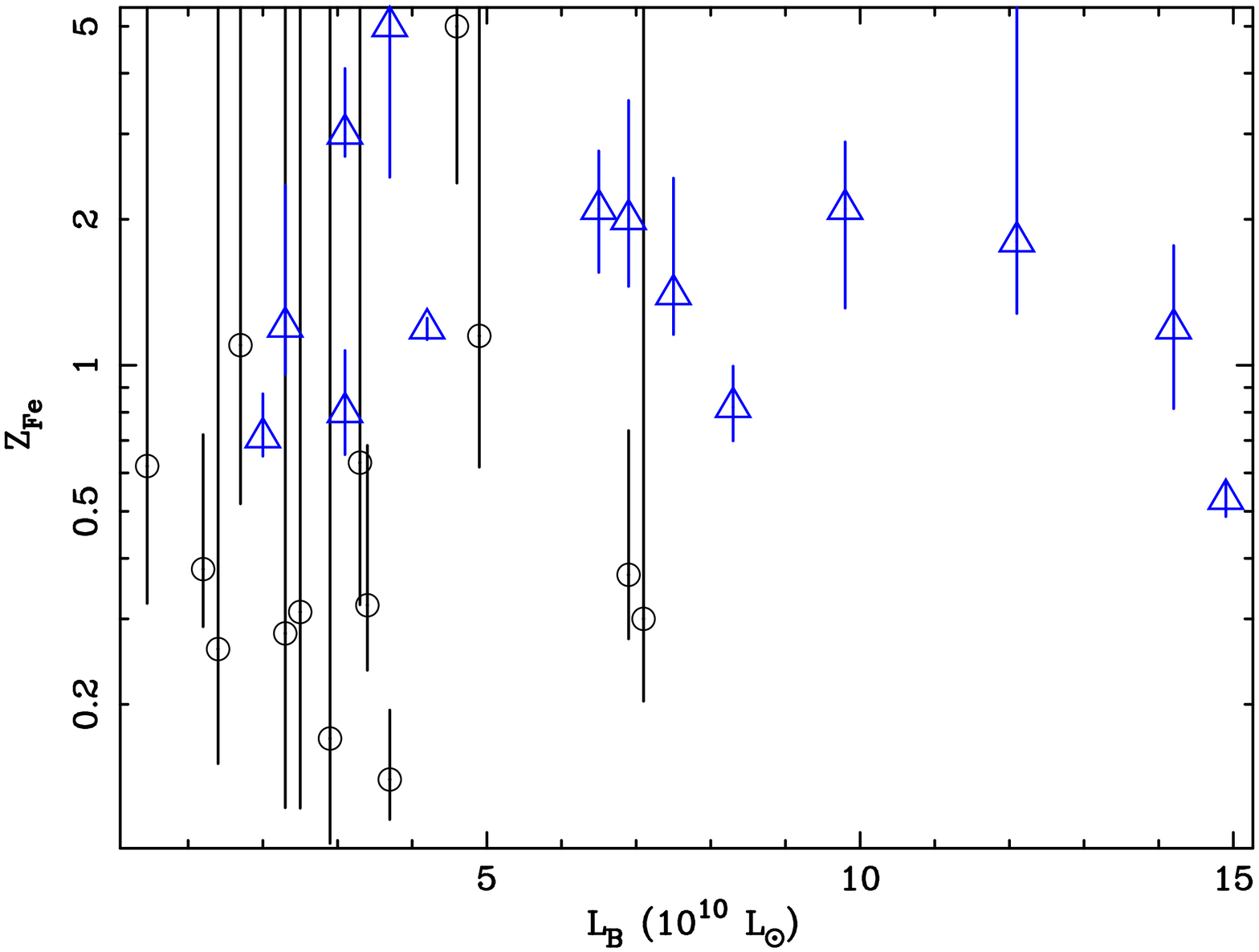}
\caption{Best-fitting \zfe\ and 1-$\sigma$ error-bars are shown as 
a function of \lx\ (left panel) and \lb\ (right panel). Data-points
marked with a circles (black) represent those galaxies for which only a single aperture
could be used, and only a single hot gas component was required by the data.
The remaining galaxies are shown as triangles (blue). \label{fig_correlation}}
\end{figure*}
\subsection{Near-solar Fe abundances} \label{sect_near_solar_abund}
In our sample of early-type galaxies, we did not find any convincing evidence 
for the very sub-solar \zfe\ historically reported for such systems.
For the brighter galaxies, we found clear evidence that the ISM abundances
are $\sim$solar. It is interesting to note that NGC\thin 507 appears to be 
something of an ``outlier'' when compared to the other high-\lx\ systems, 
since its \zfe\ seems to be somewhat lower than is typically seen in such
objects. However, at high \lx, the globally-averaged \zfe\ must start
to fall in order ultimately to match \zfe$\sim$0.4 typically seen in clusters, 
and since this is the highest-\lx\ object in our sample, its lower \zfe\
might be expected.

Although the data for the fainter galaxies are typically of
too poor quality to allow us to exclude very sub-solar abundances, they 
are generally consistent with \zfe$\sim$1. 
In fact, if we consider all those systems for which only a single hot gas
component was required in a single aperture (which are typically have 
poorer S/N), excepting the possible outlier (NGC\thin 1553; see 
\S~\ref{sect_fe_discrepancy}), which will bias the measurement due 
to its small error-bars, we obtain a mean \zfe=$0.58^{+0.27}_{-0.22}$.
To investigate any possible 
correlation between \zfe\ and \lx\ or \lb\ we applied several statistical tests---
Pearson's linear correlation test, Spearman's rank-order correlation test
and Kendal's $\tau$ test--- to the data. In Fig~\ref{fig_correlation} we show
\zfe\ plotted against \lx\ and \lb. Since it is possible that the S/N of 
the faintest galaxies masks the presence of the Fe bias, in addition to testing 
the correlation with the whole dataset we separately considered
a subset of the galaxies, in  which spatially-resolved spectroscopy or two hot 
gas models were required, thereby taking account of any possible Fe bias.
Considering all the galaxies, Pearson's linear correlation test did not 
reveal a correlation between \zfe\ and either \lx\ or \lb\ (${\rm p_0}$, 
the probability of no correlation, being 32\% and 63\%, respectively). 
With the non-parametric tests we found evidence of a correlation with 
\lx, and marginal evidence of a correlation with \lb\ 
(\eg\ for Spearman's test ${\rm p_0}$=0.5\% and 6\%, respectively). 
Considering only the ``reliable'' subset of galaxies, however, all the 
tests failed to detect a correlation (${\rm p_0 > 20}$\%), suggesting that
the correlations in the entire data-set may be artefacts related to the 
Fe bias, although this will need to be investigated with higher-quality data. 
We conclude there is no convincing evidence of a correlation between \zfe\ and
\lx\ or \lb.

The principal reason for the under-estimate of \zfe\ reported
in the literature appears to be overly simplistic spectral modelling,
as demonstrated by \citet{buote98c}. In many of the fainter systems
X-ray point-sources contribute \gtsim\ 50\% of the total
X-ray flux. Using \chandra\ we have been able to 
resolve a significant fraction of these point-sources,
reducing the impact of this ``contaminant'' 
and allowing less biased abundance determination. 
Simple fiducial models are frequently adopted to describe the 
spectral shape of the combined emission from undetected point-sources, 
but these have been by no means uniformly included in spectral-fitting
in the literature. 
It is important to understand that these are determined empirically,
and there is no {\em a priori} reason to believe them a perfect
fit to the undetected source emission in any given galaxy. This 
issue may have been, in part, responsible for the extraordinarily
low abundances reported in  three low-\lx\ galaxies observed with
\xmm\ by \citet{osullivan04a}, since the flux from the hot gas
is most likely overwhelmed completely by the point-sources, making 
accurate abundance determinations very sensitive to this modelling.

With \chandra\ we have also been able to resolve the spatially-varying
temperature structure in a number of the brighter galaxies. The 
excellent agreement between our measured abundances and previous
single-aperture \asca\ work \citep{buote98c}, where the samples overlap,
confirms that the temperature gradient is sufficient to explain 
the multiple hot gas components required in the (large) \asca\ 
extraction apertures. A similar conclusion was reached by 
\citet{buote99a} for a small number of bright galaxies, using \rosat,
which we can now confirm with the superior spectroscopic data,
and finer spatial resolution of \chandra, in a wider range of galaxies.
Resolving the spatially-varying temperature structure
mitigates the ``Fe bias'', in which the addition of multiple 
components with different temperatures tends to suppress the 
line equivalent widths, giving artificially lower abundances. 
A recent dramatic example of this effect has also been seen in the 
Antennae galaxies \citep{baldi05a}, where 
\chandra\ has revealed complex ISM structure in which \zfe\ is 
typically\gtsim1, in contrast to previous (global) \asca\ measurements of 
\zfe$\sim$0.1. 

In general, we did not require multiple hot gas components in
each annulus, indicating that the ISM is not multi-phase in these
regions.
Although, in most cases, the presence of multiple hot gas components
in the \asca\ apertures can be understood in terms of a temperature
gradient which we are able to resolve with \chandra, there still
remains the intriguing presence of multiple hot gas 
components in single annuli for several of the galaxies. In most 
cases this is only required in the innermost bin, suggesting
spatially-varying temperature structure at scales which were 
too small to accumulate useful spectra. In NGC\thin 1399 and NGC\thin 5846,
however, there was some evidence that multiple temperatures may be
needed over a range of annuli. 
In the case of NGC\thin 5846, this
simply seems to reflect the  rather disturbed morphology of the
galaxy. \citet{buote02a} first commented on the ``limited multiphase gas''
in NGC\thin 1399 based on \xmm\ data, and a similar feature is also
seen in several groups (NGC\thin 5044:\citealt{buote03a},
M\thin 87: \citealt{gastaldello02a}, RGH\thin 80:
\citealt{xue04a}). This two-temperature structure may be related
to the heating of the gas and the quenching of the cooling flow 
\citep{mathews04a}.

It is worth noting that the shapes of the  abundance profiles we
measured are consistent with observations of the centre of the 
bright group NGC\thin 5044 \citep{buote03b}, although we were
only usually able to obtain constraints for the most massive systems,
\eg\ NGC\thin 1399. Nonetheless, these abundance gradients are
similar to what are predicted by the ``circulation flow'' model
of \citet{mathews04a} (see \S~\ref{sect_enrichment_models}).

\subsection{Stellar {\em versus} ISM abundances} \label{sect_fe_discrepancy}

%\clearpage

\begin{figure*}
\centering
\plotone{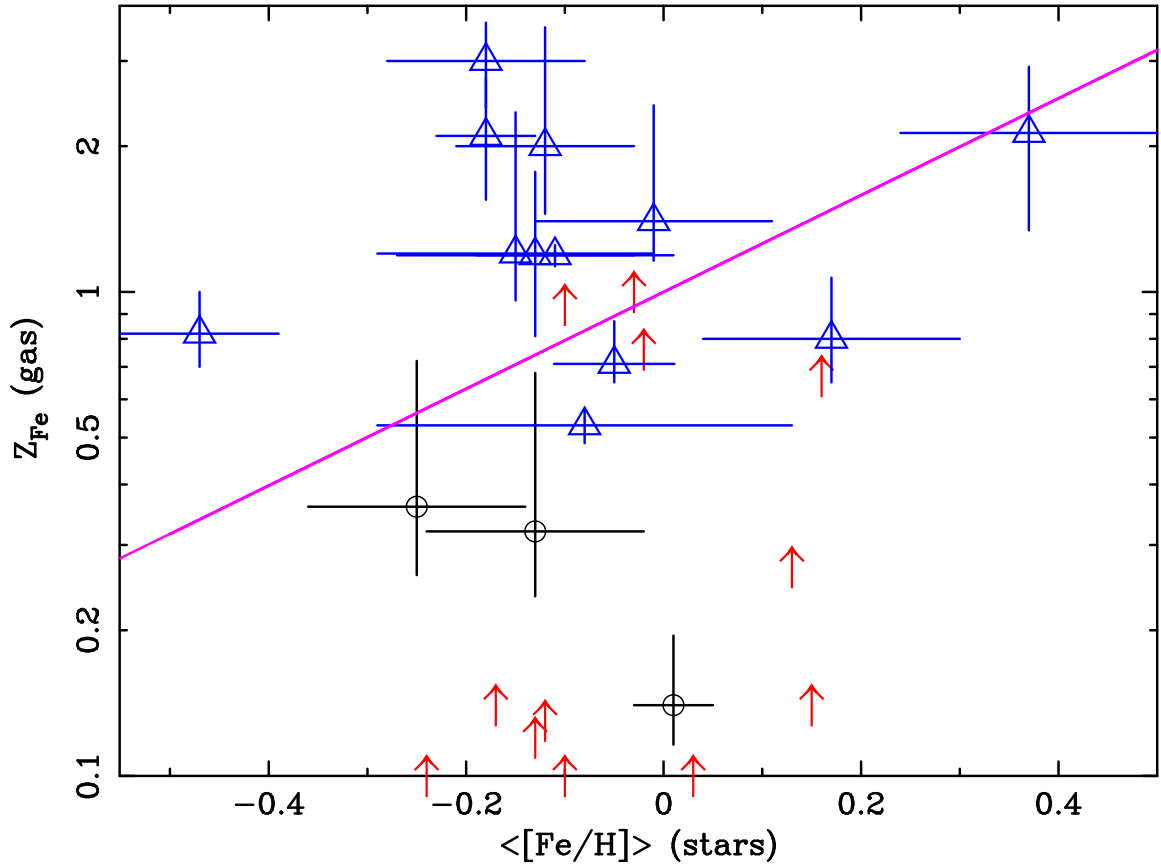}
\caption{A comparison of \zfe\ in the hot gas versus estimated
mean stellar Fe abundance. We show separately the data-points for
which spatially-resolved spectroscopy or multiple hot gas components were required 
(triangles) and those for which only a single temperature, in a single
aperture was needed (circles). 
When no upper limit
could be obtained for any galaxy, we show only the {\em lower} limit
(arrows). We show the line indicating
equal metallicity in both components. For clarity, any data-points
lying below \zfe$=0.1$ are artificially raised to this level and
all error-bars shown are 1-$\sigma$.\label{fig_abundances}}
\end{figure*}

%\clearpage

\citet{arimoto97} pointed out that the historically low ISM abundances
for early-type galaxies were in stark disagreement with the abundances
determined for the stellar content. This is difficult to interpret
in terms of simple chemical enrichment models in which the gas is 
enriched by stellar mass-loss and supernova ejecta.
In such a scenario, we might expect the ISM abundances to {\em exceed}
those in the stars
(see \S~\ref{sect_enrichment_models} for more discussion of these models).
In several galaxies we found best-fitting abundances which were 
moderately sub-solar ($\sim$0.5), and in one case 
(NGC\thin 1553) we did, in fact, obtain a best-fitting \zfe\ which
was $\sim$0.1 (although adopting a two-temperature hot gas model,
which may be more realistic in this disturbed system, the data were not
inconsistent with approximately the same ISM and stellar abundances).
To assess quantitatively whether there is still any
evidence of the discrepancy between stellar and ISM abundances
we searched the literature for optical line index measurements, from
which we were able to estimate the global Fe abundances in the stars.
We discuss this analysis in detail in Appendix~\ref{sect_stars}. 
Table~\ref{table_ssp} summarizes the properties of the stellar
population of each galaxy.

We show a comparison of the stellar and ISM \zfe\ in Fig~\ref{fig_abundances}.
The majority of the data-points  cluster 
close to the \zfe(gas)=\zfe(stars) line.
Clearly a number of points lie significantly above this line, 
suggesting that these galaxies have experienced significant
ISM enrichment. A few points lie slightly below the line, but are
approximately consistent with it within their error-bars. 
Only NGC\thin 1553 lies significantly
below the line and this galaxy has a disturbed morphology, so that
there is considerable uncertainty in using only one hot gas component
to fit the spectrum. Alternatively, the ISM abundance for this 
object may in fact be low, in which case the galaxy must have 
acquired a substantial amount of unenriched, 
primordial gas. Why such a process should seem to operate so strongly
only in this galaxy, and whether it is related to the disturbed morphology
are not entirely clear. 

If we consider those galaxies for which an upper-limit on the 
abundance could not be obtained, we find that the best-fitting
values (not shown in Fig~\ref{fig_abundances}) 
of 4 of these systems lie above and 8 below the line. This is what
we would expect if the distribution of \zfe\ in these systems is 
comparable to the ``detected'' galaxies. We therefore
conclude that these galaxies probably have a similar Fe enrichment history,
although this remains to be investigated with higher-quality data. 

Based on the \chandra\ data, therefore, we found
no evidence that the ISM abundances of the early-type galaxies
in our sample (with one exception) are significantly lower than
the mean metallicity of the stars. Furthermore, many of these
galaxies may have ISM abundances slightly higher than the stellar 
component.

\subsection{Supernova enrichment} \label{sect_sn_enrichment}

%\clearpage

\begin{deluxetable*}{llllllllll}
\tablecaption{Supernova enrichment fractions\label{table_sne}}
\tabletypesize{\scriptsize}
\tablehead{
\colhead{Galaxy} & \colhead{Elements} & \multicolumn{4}{l}{All elements} & 
\multicolumn{4}{l}{Omitting O, S} \\
\colhead{} & \colhead{} & \colhead{yield} & \colhead{$\chi^2$/dof} & \colhead{\fgas} & \colhead{${\rm f^{Ia}_{range}}$}  & \colhead{yield} & \colhead{$\chi^2$/dof} & \colhead{\fgas} & \colhead{${\rm f^{Ia}_{range}}$} 
}
\startdata
\multicolumn{10}{l}{High-\lx\ galaxies} \\ \hline
IC\thin 4296 &O,Mg,Si &WDD2 &$0.035/2$ &$0.60\pm 0.16$ &$0.28$--$1.0$ &WDD2 &$0.006/1$ &$0.60\pm 0.16$ &$0.28$--$0.63$ \\
NGC\thin 507 &O,Ne,Mg,Si,S,Ni &W7 &$16/5$ &$0.67\pm 0.05$ &$0.46$--$1.0$ &W7 &$0.95/3$ &$0.70\pm 0.05$ &$0.46$--$0.98$ \\
NGC\thin 741 &O,Mg,Si,Ni &W7 &$8.5/3.0$ &$0.88\pm 0.05$ &$0.75$--$0.91$ &W7 &$1.6/2$ &$0.70\pm 0.13$ &$0.45$--$0.94$ \\
NGC\thin 1132 &O,Mg,Si,Ni &WDD1 &$11/3$ &$0.87\pm 0.07$ &$0.65$--$0.87$ &W7 &$3.9/2$ &$0.66\pm 0.12$ &$0.43$--$0.77$ \\
NGC\thin 1399 &O,Ne,Mg,Si,S,Ni &WDD2 &$180/5$ &$0.87\pm 0.01$ &$0.70$--$0.91$ &W7 &$3.4/3$ &$0.74\pm 0.02$ &$0.49$--$1.0$ \\
NGC\thin 1600 &O,Mg,Si &W7 &$0.18/2$ &$0.68^{+0.16}_{-0.21}$ &$0.37$--$0.96$ &W7 &$0.16/1$ &$0.67^{+0.16}_{-0.21}$ &$0.37$--$0.75$ \\
NGC\thin 4472 &O,Ne,Mg,Si,S,Ni &WDD1 &$140/5$ &$0.81\pm 0.02$ &$0.53$--$0.81$ &W7 &$11/3$ &$0.60\pm 0.03$ &$0.31$--$0.66$ \\
NGC\thin 5846 &O,Ne,Mg,Si,S,Ni &WDD2 &$58/5$ &$0.88\pm 0.02$ &$0.71$--$0.92$ &W7 &$2.0/3$ &$0.73\pm 0.04$ &$0.49$--$0.87$ \\
% The old value....
%NGC\thin 7619 &O,Mg,Si &W7 &$0.48/2$ &$0.73\pm 0.11$ &$0.43$--$1.0$ &W7 &$0.46/1$ &$0.73\pm 0.11$ &$0.42$--$1.0$ \\
% The new value.... 
NGC\thin 7619 &O,Mg,Si,Ni &WDD2&$24/3$ &$0.78\pm 0.06$ &$0.49$--$0.88$ &W7&$4.4/2$ &$0.63\pm 0.07$ &$0.27$--$0.73$ \\
NGC\thin 7626 &Mg,Si &\ldots  &\ldots  &\ldots &\ldots &W7 &$0.004/1$ &$0.53\pm 0.29$ &$0.17$--$0.59$ \\
\hline \multicolumn{10}{l}{Moderate-\lx\ galaxies} \\ \hline
NGC\thin 720$^1$ &O,Ne,Mg &WDD1 &$21/2$ &$0.85\pm 0.06$ &$0.63$--$0.85$ &WDD1 &$2.4/1$ &$0.58\pm 0.12$ &$0.02$--$0.58$ \\
NGC\thin 1332$^1$ &O,Ne,Mg,Si &WDD2 &$37/3$ &$0.72\pm 0.05$ &$0.39$--$0.75$ &W7 &$2.7/2$ &$0.55\pm 0.07$ &$0.16$--$0.56$ \\
NGC\thin 1387 &O,Mg,Si &W7 &$0.003/2$ &$0.94^{+0.06}_{-0.18}$ &$0.82$--$0.94$ &W7 &$0.002/1$ &$0.94^{+0.06}_{-0.18}$ &$0.82$--$0.94$ \\
NGC\thin 1407 &O,Mg,Si,S,Ni &WDD1 &$39/4$ &$0.84\pm 0.04$ &$0.61$--$0.84$ &W7 &$1.8/2$ &$0.58\pm 0.08$ &$0.24$--$0.62$ \\
NGC\thin 1553 &O,Ne &WDD2 &$13/1$ &$0.71\pm 0.09$ &$0.33$--$0.71$ &WDD1 &$0/0$ &$0.42\pm 0.16$ &$0.0$--$0.43$ \\
NGC\thin 1700 &O,Ne,Mg &WDD2 &$3.2/2$ &$0.84\pm 0.08$ &$0.65$--$0.84$ &WDD1 &$2.3\times 10^{-7}/1$ &$0.76\pm 0.11$ &$0.49$--$0.76$ \\
NGC\thin 3607 &Mg &\ldots &\ldots  &\ldots &\ldots &W7 &$0/0$ &$0.76\pm 0.24$ &$0.33$--$0.77$ \\
NGC\thin 3923 &O,Ne,Mg,Si &WDD1 &$39.1/3$ &$0.89\pm 0.05$ &$0.71$--$0.97$ &WDD1 &$5.4/2$ &$0.56\pm 0.11$ &$0$--$0.56$ \\
% The old value....
%NGC\thin 4365 &O,Mg &W7 &$0.45/1$ &$0.60\pm 0.23$ &$0$--$0.60$ &W7 &$0/0$ &$0.70\pm 0.30$ &$0.16$--$0.70$ \\
% The new value....
NGC 4365 &O,Mg &WDD1&$0.48/1$ &$0.77^{+0.19}_{-0.27}$ &$0.50$--$0.77$ &WDD2&$0/0$ &$0.67\pm 0.34$ &$0.076$--$0.67$ \\
NGC\thin 4552 &O,Ne,Mg,Si,Ni &WDD1 &$76/4$ &$0.89\pm 0.02$ &$0.73$--$0.92$ &W7 &$3.1/3$ &$0.73\pm 0.04$ &$0.42$--$0.74$ \\
\enddata
\tablecomments{The inferred SNIa enrichment fraction, \fgas, 
for the ISM of the sample galaxies. The preferred SNIa metal yield model is 
listed for each galaxy (``yield''), whereas we adopt the N97 metal yields
for SNII. The range of best-fitting \fgas values,
${\rm f^{Ia}_{range}}$, for different combinations of SNII and SNIa yields 
(see text) are also shown. We show the results derived from all the 
measured abundance ratios (``All elements'') and those which omit,
if possible, O and S, both of which seemed aberrant (see text).}
\end{deluxetable*}

%\clearpage
With our \chandra\ data we have been able to obtain, in many cases for 
the first time, interesting constraints on $\alpha$-to-Fe abundance ratios.
A number of competing effects can enrich or dilute the ISM metal content 
in early-type galaxies, including supernovae ejecta, 
stellar winds and large-scale galactic gas inflow/ outflow
\citep{matteucci86}. The $\alpha$-elements and Fe are predominantly
processed through supernovae, so stellar mass-loss acts only as a 
``slow-release reservoir'' for the elements; if the stars were all formed 
very quickly, we would expect the metal yields of the 
mass-losing stars to resemble SNII. Because these element are
primarily produced in supernovae it is common-practice to define, 
as a fiducial diagnostic, the fraction of the Fe content of a 
galaxy produced in SNIa, f. Since Fe is more profusely synthesized in 
SNIa than SNII, typically this quantity is measured 
by matching linear combinations the SNe yields for a complete
stellar population (integrated over an assumed IMF; typically that of
Salpeter) to the abundance ratios with respect to Fe
\citep[\eg][]{gastaldello02a}.

It is important to stress that, on account of the 
finite timescale over which the gas is processed through stars
the SNIa fraction measured in the hot gas, \fgas\ and 
that determined from the stars, \fstars,  are not identical,
and nor are they entirely commensurate with the total fraction
of Fe synthesized in SNIa during the galaxy history
\citep{matteucci04a}. 
However \fgas\ and \fstars\
can be calibrated against realistic gasdynamical enrichment 
simulations \citep[\eg][]{brighenti99a}, making them powerful measures of 
the enrichment history of a galaxy; in any case, they provide a 
useful  yard-stick with which to compare different galaxies.
Furthermore the  relative abundances of elements which are 
predominantly products of one type of supernova
are fixed by the SNe yields, since there is unlikely to be significant
segregation of metals produced together. This enables measurement of 
the abundance ratios to provide direct constraints on the 
SNe yields and, in principle, the IMF. 
Given the remarkable degree of scatter in the predicted
metal yields between different explosion models 
\citep[][hereafter G97]{gibson97},
such constraints are of particular importance.

In order to determine values of \fgas\ for each galaxy
we performed fitting analogous to that of \citet{gastaldello02a}.
We experimented with different SNe metal yields, which are highly
sensitive to the supernova modelling. We adopted
the SNIa W7, WDD1 or WDD2 yields from \citet{nomoto97b}
and, for ease of comparison with other authors, the 
SNII yields from \citet[][hereafter N97]{nomoto97a}. We also 
experimented with a range of SNII yields taken from G97, 
in order to examine the sensitivity
of the results to the adopted model. In practice the G97 yields
are less useful than N97 since they do not all include the same
range of species. It is worth noting that all of these models assume
a Salpeter IMF, although the IMF is not expected to affect the 
abundance ratios as significantly as the overall abundances (G97).
We note that fitting this model to the \zo/\zfe, \zne/\zfe,
\zmg/\zfe, \zsi/\zfe, \zs/\zfe and \zni/\zfe\ ratios observed in the Sun
\citep{asplund04a} yields the well-known \fstars$\sim$0.75 appropriate
for the Solar neighbourhood whilst strongly favouring the WDD2 SNIa 
yields.

Initially we fitted our model to  all of the available $\alpha$/Fe 
abundance ratios, adopting the SNIa model which best fitted the data.
There is remarkable agreement between the supernova fractions measured
in each galaxy, with \fgas${\sim}$ ranging from 0.7--0.9, although
it fell as low as $\sim$0.3 if the G97 yields are 
used.  The results are summarized in Table~\ref{table_sne}. 
Such values are in excellent agreement with recent observations of the 
centres of groups and clusters. For example \fgas$\sim$0.8--0.9 
in M\thin 87 \citep{gastaldello02a}, \fgas$\sim$0.7--0.8 in
NGC\thin 5044 \citep{buote03b} and \fgas$\sim$0.8--0.9 in the centre
of the cluster A\thin 1795 \citep{ettori02a}.
We found a general trend that the delayed detonation SNIa yields
were preferred over the deflagration (W7) model, although this 
seems to be driven by the low \zo/\zmg\ ratios in these models,
which could be a source of uncertainty (see below).

In contrast to the enrichment of the gas, super-solar $\alpha/Fe$ ratios 
are typically reported  in the stars when simple stellar population
models are fitted (see Appendix~\ref{sect_stars}), which implies substantially
more Type~II enrichment in the stars. 
Assuming the $\alpha/Fe$ abundances obtained in Table~\ref{table_ssp} 
are driven primarily by the Mg/Fe ratio (and assuming that there is 
no stellar Mg/Fe gradient), we estimate an average
\fstars${\sim 0.35}$, with individual values ranging from
$\sim 0.60$ for NGC\thin 5018 to $\sim 0.1$ for NGC\thin 1399, adopting
the W7 SNIa and N97 SNII yields. This is a well-known result 
\citep[\eg][]{worthey98} which is typically explained in terms of the short 
timescale of star-formation in early-type galaxies, or variations in the 
IMF. The stark discrepancy between \fgas\ and \fstars\
can readily be explained only if substantial amounts of Fe are added
to the ISM through SNIa occurring after the bulk of the stars have formed.
Such a large input of Fe into the ISM would tend to raise \zfe\ to
$\sim$2--3 times the mean stellar value. The error-bars
on our measured \zfe\ in the gas are sufficiently large that many of 
the galaxies in the sample could have \zfe 
consistent with such an elevated value;
in particular we note that NGC\thin 1332 and NGC\thin 4472 have
best-fit abundances close to that required by this argument. 
However, several galaxies (most notably NGC\thin 1399 and 
NGC\thin 720) seem to have \zfe\ about a factor $\sim$2 too low,
implying dilution of the ISM by a source of primordial 
(\ie\ un-enriched) gas in these systems.

Despite consistency with \fgas\ determined in other systems, 
it is clear that the fits are typically 
rather poor, suggesting a disagreement between the standard SNe yields
and the ISM abundance pattern. This primarily arises because
of discrepant $\alpha$-element abundance ratios, 
even taking into account the large scatter in the SNII yields.
The chiefly discrepant species is O, which appears to be significantly
under-abundant in comparison to other $\alpha$-elements, such as Mg
(for which we have the most measurements). The measurement of reliable 
\zo\ in the X-ray band is challenging, especially when 
attempting imaging spectroscopy 
at energies where the CCD  response may not be perfectly well understood 
(\S~\ref{sect_systematics}). At plasma temperatures $\sim$0.5--1.0~keV,
appropriate for our sample, the principal O features tend to be 
blended somewhat with the Fe L-shell ``hump'' which exacerbates the 
difficulty in reliably determining \zo. 
However, unexpectedly low values of 
\zo/\zmg\ ($\sim$0.8--1.3, given our default solar abundances) 
have been widely reported in the literature, in contrast to 
\zo/\zmg $\sim$1.4--2.8 predicted by SNII models (G97).
This discrepancy  has been found in both imaging and grating spectroscopy
data from a variety of different instruments observing  a wide range of 
systems such 
as the bright group NGC\th 5044 \citep{buote03b,tamura03a},
M\th 87 \citep{gastaldello02a,sakelliou02}, the massive elliptical
NGC\thin 4636 \citep{xu02a}, the starbursting galaxy M\th 82 
\citep{tsuru97} and the centres of some galaxy clusters 
\citep{peterson03a}. There is even some evidence of this effect
in the Milky Way ISM \citep{ueda05a}.

It is not certain whether the observed O deficiency
represents a problem with the SN yields,
which are highly sensitive to the (rather uncertain) SNe modelling
(G97), or if it is indicative of an additional source of metal 
enrichment within
the galaxy. It is interesting to note that some SNR have 
been observed with apparent abundance anomalies (for example 
N49B in which the ejecta appear significantly enriched in Mg without
accompanying Ne or O: \citealt{park03}), which might hint at a 
problem with the nuclear physics. In contrast,
\citet{loewenstein01} suggested there may be an alternative
source of metal enrichment to explain
the low \zo/\zfe\ in clusters. He proposed that 
significant enrichment in the early universe may arise from population III
hypernovae, since the O-burning region is significantly expanded in an hypernova. 
Hypernova metal yields have been computed by \citet{umeda02a} and
\citet{heger02a}, but unfortunately they did not exhibit 
the very low \zo/\zmg\ (nor \zo/\zne) which are observed in our data, 
although the 
calculations for Mg and Ne production are  sensitive to the model
assumptions. Given the published Pop~III hypernova yields
\citet{matteucci05a} pointed out they would not have a significant 
impact upon the abundance pattern of an elliptical galaxy.

Given these reservations, we did not attempt to include a
putative hypernova component in our modelling. If, instead, the discrepancy 
between the observed and predicted abundance of O reflects 
a problem in the computed yields, notwithstanding their success in 
reproducing the Solar abundance pattern, an alternative approach is 
simply to omit this element from the computation of \fgas. This tended
to result in slightly lower estimated values of \fgas, but substantially
improved $\chi^2$. Interestingly
we also found that \zs/\zfe, where it could be measured, was slightly 
under-predicted by the model (which may be a systematic error;
\S~\ref{sect_systematics_calibration}) and so we also omitted this 
element, 
although it had little impact on the best-fitting \fgas.
The fit results are shown in Table~\ref{table_sne}. 
We found a mean \fgas=$0.66\pm0.11$, where the quoted 
error is the standard deviation in the measurements.
%There is little evidence of any trend with galaxy properties. 
Using the same statistical tests described in \S~\ref{sect_near_solar_abund}
we searched for correlations between \fgas\ and \lx, \lb\ and 
\zfe, but found no evidence of any such correlations 
(${\rm p_0 > 40}$\% in all cases).
In this case, we found that the 
W7 (deflagration model) abundance yields are preferred over the 
delayed detonation models, although in most cases (with one
or two notable exceptions, such as NGC\thin 1399) the data do
not strongly discriminate between them.
It is striking that \fgas\ in these galaxies is remarkably 
similar to that in the 
Solar neighbourhood, implying a similar enrichment history
for the hot ISM of early-type galaxies and the cold ISM of
spirals.

An intriguing piece of evidence which further complicates this 
picture of metal enrichment is the observation that [O/Mg] in Galactic
bulge stars appears to be a decreasing function of metallicity
\citep{fulbright04a}. 
There is a reduction in [O/Mg] by  as much as $\sim$0.5~dex with
respect to the SN~II yield level by \zfe$\sim$1, providing 
evidence that O may be under-abundant in old stellar populations, not
just the ISM. Moreover such a trend is very difficult to reconcile
with a picture where only SNII and SNIa enrich the gas. Nonetheless
whether a putative Pop~III hypernova population could resolve this issue 
is unclear.

\subsection{Enrichment models} \label{sect_enrichment_models}
We now discuss our results in the light of various models for
the chemical enrichment of early-type galaxies. In traditional ``galactic
wind models'' the galaxies form monolithically and passively
evolve, during which time the dynamics and luminosity 
of the gas is strongly affected by supernova feedback, which can drive
periods of large scale galactic outflow \citep[\eg][]{ciotti91}. 
Given the large amount of metals injected into the ISM by the 
supernovae, these models tend to predict \zfe$\sim$a few times solar, 
(and, certainly, dramatically in excess of the abundance of the stars),
which is clearly in disagreement with the majority of galaxies in 
the sample, in many cases by as much as an order of magnitude.
\citet{pipino05a} allowed the accretion of additional, unenriched
gas, but were unable to mitigate this discrepancy while reproducing
the other optical and X-ray properties of the gas. 
In contrast, \citet{brighenti99a} constructed a cooling-flow gasdynamical model
for NGC\thin 4472, incorporating the prolonged infall of gas. These
authors were able to produce $\sim$0.5 solar abundances for the galaxy
(the precise abundances being rather sensitive to the assumed supernovae
rates), although the density profile was too peaked on account of 
excessive cooling in the core. Furthermore, the enrichment history
assumed in this model makes it difficult to understand the large
amount of metals in clusters in terms of the contribution from 
elliptical galaxies.

An alternative approach was adopted by \citet{kawata03a}, who followed
the evolution of a $\sim 10^{13}$\msun\ elliptical galaxy in an
hierarchical structure formation simulation. In their model, however, 
most of the  Fe produced by the stars is locked into large amounts of cool 
($\sim 10^4$~K) gas and excessive on-going star formation, 
at variance with observations.
In this case the hot phase ISM was found to have 
\zfe\ltsim 0.3 and, critically \zfe(stars) was
$\sim$an order of magnitude greater than that of the hot gas.
The results are clearly at variance with our \chandra\ measurements,
in the opposite sense to the galactic wind models. 

One key ingredient not typically incorporated into enrichment models
is the role of AGN feedback. There is some evidence of disturbances
in some ``normal'' early-type galaxies which may point to 
interaction between the ubiquitous supermassive central black-hole
and the ISM \citep[\eg][]{blanton01b,jones02a}.
Numerical simulations have suggested that
ISM-AGN interactions can have a dramatic effect on the 
dynamics and energetics of the ISM \citep{dimatteo05a}.
\citet{kawata05a} incorporated AGN feedback into their model,
finding that this effectively evacuated
the (cool) enriched gas from the galaxy, so that the discrepancy
between stellar and ISM abundances of their model persists.
We note that these authors comment on the good agreement between
their model and \zfe\ for the galaxy NGC\thin 3923, based on the results of 
\citet{matsushita00a}. Our analysis of the \chandra\ data indicates
a significantly higher abundance for this galaxy, underlining the
discrepancy between our results and 
the predictions of this model.
\citet{mathews04a}, in contrast, showed that allowing a small 
amount of continuous heating from a central AGN may drive 
large-scale ``circulation flows'', which efficiently
mixes gas through the galaxy. This model could reproduce
the abundance gradients found in the \xmm\ and \chandra\ data of the 
group NGC\thin 5044. Crucially it predicted the correct \zfe, 
which is coincidentally similar to that observed in most galaxies
in our sample. Whether a variant of this model
can successfully describe the lower-mass systems studied in this
present work remains to be seen, however.

\section{Summary}
We have surveyed the metal abundances of the hot phase ISM within a 
sample of 28 early-type galaxies, using \chandra. In summary:
\begin{enumerate}
\item We did not find the historically very sub-solar 
abundances reported for early-type galaxies, nor is there convincing
evidence that \zfe\ correlates with the galaxy luminosity. Considering
just our lowest-S/N galaxies, we found a mean \zfe=0.58$^{+0.27}_{-0.22}$.
The discrepancy
with past results
appears to be due to imperfect modelling of the (undetected) 
point-source populations (of which \chandra\ has enabled a 
substantial fraction to be detected) and the fitting
of single-temperature models to multi-temperature gas (the Fe
bias).
\item We confirmed the conclusions of \citet{buote98c} that multiple
components are often required to fit the spectra of early-type galaxies.
In a number of cases the excellent spatial resolution of \chandra\
enabled us to resolve temperature gradients lying within \asca\
or \rosat\ spectral extraction apertures. In a few
cases the fit required multiple hot gas components in a given annulus. In 
all cases a component was required to account for undetected 
point-sources.
\item Comparing the emission-weighted ISM and stellar Fe abundances
we found excellent agreement. Excepting one possible outlier (NGC\thin 1553)
there is no evidence that the stellar abundances are significantly
higher than those of the gas. In fact the data allow the ISM 
\zfe\ to be slightly higher on average.
\item The data are inconsistent with both monolithic ``galactic
wind models'', which over-predict \zfe\ and the recent
hierarchical models of \citet{kawata05a}, which under-predict
\zfe. The ``circulation flow'' model of \citet{mathews04a}
is able to explain \zfe\ of about the correct magnitude, albeit for a system
more massive than most considered here.
\item We found that the SNIa Fe enrichment fraction in the hot gas, \fgas,
is $66\pm11$\% and shows no obvious trend with luminosity.
This value is remarkably close to the 75\% enrichment fraction for
the solar neighbourhood, indicating similarity in the enrichment
histories for the early-type and spiral galaxies. It is also
similar to that observed in the centres of clusters and groups,
indicating homology of enrichment from cluster scales to 
intermediate-\lx\ galaxies.
\item The \zmg/\zfe\ ratio in the gas is significantly lower than
that in the stars, indicating that significant SNIa must enrich the
gas, in addition to stellar mass-loss. In at least two cases
(NGC\thin 1399 and NGC\thin 720) this would seem to require the
further accretion of unenriched, ``primordial'' material to prevent
\zfe\ being significantly higher than observed.
\item There is increasing evidence that O is over-predicted by 
the SNII yields, which may imply problems in the supernova 
calculations.  Alternatively, the measurement of varying 
[O/Mg] in Galactic bulge stars may point to a source of 
$\alpha$-element enrichment in addition to SNIa and SNII.
\end{enumerate}

\appendix
\section{Stellar abundance estimates} \label{sect_stars}
In order to estimate the mean Fe abundance  in the stars, we searched the 
literature to obtain reliable Lick/IDS index \citep{faber85} 
measurements for the centre of  each galaxy. In order to break the age-metallicity 
degeneracy, we converted the line indices into \feh\ by linearly 
interpolating the simple stellar population (SSP) model results of 
\citet{thomas03a}. These models take into account non-solar $\alpha$-element to Fe 
ratios, which are typically reported in early-type galaxies
\citep[\eg][]{trager00a}. Where possible, we fitted the interpolated model
with a $\chi^2$ minimization procedure to the central 
H$\beta$, Mgb, Fe5270 and Fe5335 indices. In several cases some or all of these
indices were not available in the literature, in which case we 
either adopted appropriate alternative indices or omitted them.
In a number of cases one or more of the indices appeared to be discrepant
with the others, which may be a consequence of systematic measurement
errors (or, possibly, the breakdown of the SSP model assumptions). 
In these cases, we omitted the most discrepant index.
In Table~\ref{table_ssp} we list the indices used for each galaxy. 
For NGC\thin 1387, we could not find useful line index measurements in the 
literature, and so we omitted it from the comparison with the X-ray
results (\S~\ref{sect_fe_discrepancy}).
For several galaxies, it was necessary to fix the age or [$\alpha$/Fe]
in order to obtain interesting abundance constraints. In these cases,
we adopted an age of 12~Gyr and [$\alpha$/Fe]=0.23, which are 
roughly consistent with the average values found in the remaining galaxies.
The best-fitting parameter values, converted to the solar 
abundances standard of \citet{asplund04a}, 
and the reference from which the line strength measurements were derived are 
shown in Table~\ref{table_ssp}.

It is interesting to compare our results with those of other recent 
catalogues of ages and metal abundances in early-type galaxies
which overlap with our sample.
\citet{thomas05a} measured ages 
and central abundances for a sample of galaxies, of which 14 overlapped our
sample, using the SSP models adopted in this present work. 
In general we found good agreement with our results; the only two
galaxies for which \zfe\ was substantially discrepant
were NGC\thin 1600 and NGC\thin 7619, for which our abundances 
were $\sim$0.1~dex higher. 
\citet{terlevich02} and  \citet{trager00a} both estimated (central)
metal abundances, ages and $\alpha$/Fe ratios using the older SSP models
of \citet{worthey94a}.
\citet{trager00a} made corrections for 
non-solar $\alpha$/Fe ratios so that [Fe/H] could explicitly be obtained.
In contrast, \citet{terlevich02} adopted the [MgFe] lick index, which is more
sensitive to the overall abundance (${\rm [Z/H]=[Fe/H]+0.94[\alpha/Fe]}$ for the 
models we adopt).
In general our results agree reasonably well ($\pm 2\sigma$) with these 
previous measurements
where the data overlap. There was some scatter,
however; for example we found that our central [Fe/H] were 
typically $\sim$0.1~dex higher 
than those found by \citet{trager00a}. We found the most 
significant discrepancies (\gtsim 3$\sigma$) between our results and those of 
\citet{terlevich02} for the galaxies NGC\thin 1553, NGC\thin 3115 and NGC\thin 3607, 
for all of which those authors reported poor quality data. Given the 
good agreement with \citet{thomas05a}, the differences are likely a consequence
of the different SSP models used.

To estimate the total Fe content of the stars it is necessary to take
into account any abundance gradients. There is a substantial body of 
literature on line index gradients in early-type galaxies
\citep[see][and references therein]{kobayashi99a}. 
Probably the most extensive collection of abundance gradient
measurements in early-type galaxies overlapping our sample is that of
\citet{kobayashi99a} who used linear
approximations for the line-strength--metallicity relations at fixed
galaxy ages and treated each line index  separately.
Unfortunately, their results were very sensitive to the line index 
adopted, and their method for relaxing the age constraint was not 
always successful for each galaxy (see their results for NGC\thin 720).
A fully self-consistent assessment of stellar abundance gradients in all of 
our galaxies is beyond the scope of this present work.
However, for those galaxies in the sample of \citet{trager00a}, we fitted
simultaneously the data in both the ${\rm r_e/8}$ and ${\rm r_e/2}$
apertures
(where ${\rm r_e}$ is the effective radius of the galaxy), 
assuming that the age and $\alpha$/Fe ratios are the same
in each aperture. This allowed us to estimate the magnitude of the stellar
abundance gradient. We note that \citet{trager00a} allowed for both age
and $\alpha$/Fe ratio gradients, but we typically did not find this necessary
to obtain good fits. In any case, if the overall stellar population 
comprises a  mixture of different-aged components, the application of 
single burst SSP models is probably overly simplistic. 
To estimate the global emission-weighted Fe abundance, we
adopted the prescription of \citet{arimoto97}, who assumed 
\zfe${\rm \propto r^{-c}}$ for the stars, where r is the distance from the
galactic centre. Assuming a circular aperture, which is appropriate
for these line index measurements \citep[see][]{gonzales93a},
and a de Vaucouleurs optical brightness profile,
we can express the emission weighted abundance within aperture r as:
\[
\bar{{Z}_{Fe}}(r) = \bar{{Z}_{Fe}}(\infty) \frac{P(8-4c,7.67(r/r_e)^{1/4})}{P(8,7.67(r/r_e)^{1/4})}
\]
where $P(a,x)$ is the incomplete gamma function
$\int_0^x dt\  \exp(-t) t^{a-1}/\Gamma(a)$.
Two abundance measurements in 
different apertures are therefore sufficient to constrain both unknowns,
c and $\bar{{Z}_{Fe}}(\infty)$, of which the latter is the quantity of interest for 
comparison with the X-ray results. Following convention, we henceforth denote
${\rm \log_{10}(\bar{Z_{Fe}}(r_e/8))\equiv[Fe/H]_0}$ and 
${\rm \log_{10}(\bar{Z_{Fe}}(\infty))\equiv<[Fe/H]>}$. For the 10 of our galaxies
in the \citet{trager00a} sample, we found that the weighted mean of the 
difference between ${\rm <[Fe/H]>}$  and [Fe/H]$_0$ was $-0.27\pm0.06$. 
This corresponds to c=0.24, in good agreement with the average
$\sim$0.3 found by \citet{kobayashi99a}. Therefore, we used this factor
to estimate the global abundances where we only had central line indices.
The assumed
power law relation for \zfe\ must break down at small radii,
since it inflates unphysically, and so, for convenience, we assume
that any ``central'' abundances inferred from the literature are the same as
would be measured in an ${\rm r_e/8}$ aperture. In practice correcting from
an aperture of size ${\rm r_e/10}$ to one of size ${\rm r_e/8}$ 
involves shifting the abundances
only by 0.02~dex, which is considerably smaller than the statistical errors.

%\clearpage

\begin{deluxetable*}{lllrrrr}
\tablecaption{Stellar population parameters\label{table_ssp}}
\tabletypesize{\footnotesize}
\tablehead{
\colhead{Galaxy} & \colhead{indices} & \colhead{ref.} & \colhead{${\rm [\alpha/Fe]}$} & \colhead{age} & \colhead{${\rm [Fe/H]_0}$} & \colhead{$\rm <[Fe/H]>$} \\
\colhead{} & \colhead{} & \colhead{} & \colhead{} & \colhead{(Gyr)} & \colhead{} & \colhead{} }
\startdata
IC\thin 4296  & Mg2, Mgb, Fe5270, Fe5335       & 2 & 0.31$\pm0.08$ & 12 & 0.12$\pm$0.10 & -0.10$\pm$0.11\ddag\\
NGC\thin 507  &  H$\beta$, Mgb, Fe5270, Fe5335 & 4 & 0.30$\pm0.05$ & 7.0$\pm3.0$ & 0.00$\pm$0.11 & -0.08$\pm$0.35 \\
NGC\thin720   & H$\beta$, Mgb, Fe5270, Fe5335  & 4 & 0.37$\pm0.05$ & 2.9$^{+1.3}_{-0.3}$ & 0.30$\pm$0.14& 0.17$\pm$0.21 \\
NGC\thin 741 & H$\beta$, Mgb, Fe5270, Fe5335   & 5 & 0.20$\pm0.16$ & 12 & 0.09$\pm$0.22 & -0.13$\pm$0.23\ddag\\
NGC\thin 1132 & Mg2                            & 3 & 0.23 & 12 & -0.25$\pm$0.11 & -0.47$\pm$0.13\ddag \\
NGC\thin 1332 & H$\beta$, Mgb, Fe5270, Fe5335  & 5 & 0.31$\pm0.16$ & 12 & 0.07$\pm$0.22 & -0.15$\pm$0.23\ddag \\
NGC\thin 1399 &  H$\beta$, Mgb, Fe5270, Fe5335 & 1 & 0.37$\pm0.05$ & 12$^{+4}_{-3}$ & 0.11$\pm0.11$ & -0.11$\pm0.13$\ddag \\
NGC\thin 1407 &  H$\beta$, Mgb, Fe5270, Fe5335 & 1 & 0.33$\pm 0.02$ & 12$\pm2$ & 0.04$\pm0.06$ & -0.18$\pm0.08$\ddag\\
NGC\thin 1549 &  H$\beta$, Mgb, Fe5270, Fe5335 & 1 & 0.24$\pm0.01$ & 5.1$\pm0.7$ & 0.12$\pm0.03$ & -0.10$\pm0.07$\ddag \\
NGC\thin 1553 &  H$\beta$, Mgb, Fe5270, Fe5335 & 1 & 0.17$\pm0.01$ & 5.7$^{+1.2}_{-0.8}$ & 0.23$\pm$0.04 & 0.01$\pm$0.07\ddag \\
NGC\thin 1600 & H$\beta$, Mgb, Fe5335          & 4 & 0.24$\pm0.02$ & 6.6$\pm2.3$ & 0.30$\pm$0.07 & 0.37$\pm0.21$ \\
NGC\thin 1700 &  H$\beta$, Mgb, Fe5270, Fe5335 & 4 & 0.16$\pm0.02$ & 2.8$^{+0.5}_{-0.4}$ & 0.31$\pm$0.08 & 0.16$\pm$0.12\\
NGC\thin 3115 & H$\beta$, Mgb, Fe5270, Fe5335  & 5 & 0.11$\pm0.11$ & 12 & 0.25$\pm0.16$ & 0.03$\pm0.17$\ddag\\
NGC\thin 3585 & H$\beta$, Mgb, Fe5270, Fe5335  & 5 & 0.20$\pm0.12$ & 12 & 0.09$\pm$0.17 & -0.13$\pm$0.18\ddag\\
NGC\thin 3607 & H$\beta$, Mgb, Fe5270, Fe5335  & 5 & 0.19$\pm0.12$ & 15$\pm4$ & 0.09$\pm$0.17 & -0.13$\pm$0.18\ddag\\
NGC\thin 3608 &  H$\beta$, Mgb, Fe5270, Fe5335 & 4 & 0.19$\pm0.03$ & 8.2$\pm2.0$ & 0.16$\pm$0.07 & -0.17$\pm$0.16 \\
NGC\thin 3923 &  H$\beta$, Mgb, Fe5270, Fe5335 & 1 & 0.34$\pm$0.03 & 3.3$^{+0.7}_{-0.3}$ & 0.35$\pm$0.08 & 0.13$\pm$0.10\ddag \\
NGC\thin 4365 &  H$\beta$, Mgb, Fe5270, Fe5335 & 5 & 0.19$\pm$0.14 & 12 & 0.20$\pm$0.18 & -0.02$\pm$0.19\ddag\\
NGC\thin 4472 &  H$\beta$, Mgb, Fe5270, Fe5335 & 4 & 0.25$\pm0.03$ & 9.0$\pm2.0$ & 0.13$\pm0.06$ & -0.01$\pm0.20$ \\
NGC\thin 4494 &  H$\beta$, Mgb, Fe5270, Fe5335 & 5 & 0.16$\pm0.13$ & 12 & -0.02$\pm$0.19 & -0.24$\pm$0.19\ddag\\
NGC\thin 4552 &  H$\beta$, Mgb, Fe5335         & 4 & 0.24$\pm0.02$ & 12$\pm1.7$ &  0.17$\pm0.04$ & -0.05$\pm$0.10 \\
NGC\thin 4621 &  H$\beta$, Mgb, Fe5335         & 5 & 0.28$\pm$0.14 & 12 & 0.19$\pm$0.17 & -0.03$\pm$0.18\ddag\\
NGC\thin 5018 &  H$\beta$, Mgb, Fe5335         & 5 & 0.01$\pm$0.14 & 2.0$^{+2.3}_{-0.4}$ & 0.37$\pm$0.27 & 0.15$\pm$0.28\ddag\\
NGC\thin 5845 &  H$\beta$, Mgb, Fe5335         & 5 & 0.26$\pm$0.17 & 12 & 0.10$\pm$0.25 & -0.12$\pm$0.26\ddag\\
NGC\thin 5846 & H$\beta$, Mgb, Fe5335          & 4 & 0.22$\pm0.02$ & 15$\pm$2.0 &  0.05$\pm0.04$ & -0.18$\pm0.16$ \\
NGC\thin 7619 & H$\beta$, Mgb, Fe5335          & 4 & 0.18$\pm0.02$ & 13.5$\pm1.0$& 0.21$\pm0.04$ & -0.12$\pm0.15$ \\
NGC\thin 7626 & H$\beta$, Mgb, Fe5335          & 4 & 0.24$\pm0.02$ & 14$\pm2$ & 0.08$\pm0.05$ & -0.25$\pm0.18$ \\
\enddata
\tablecomments{The mean abundances determined from fitting SSP models to the 
Lick indices reported in the literature. 
Those mean stellar abundances (${\rm <[Fe/H]>}$) marked 
\ddag\ were estimated from the central abundance (${\rm [Fe/H]_0}$) 
assuming a constant abundance gradient; otherwise the mean
abundance is inferred by extrapolation (see text). Where no error-bar
is given, the parameter was frozen.
Table references: 1--- \citet{beuing02a}, 2--- \citet{gorgas90},
3--- \leda\, 4--- \citet{trager00a}, 5--- \citet{trager98a} }
\end{deluxetable*}

%\clearpage

\begin{acknowledgements}
We would like to thank Fabio Gastaldello for useful discussions
concerning data analysis and interpretation, and for advice
and support with the \xmm\ data-reduction.
We would also like to thank Bill Mathews and Fabrizio Brighenti for 
stimulating discussions on the enrichment processes in early-type 
galaxies. 
This research has made use of data obtained from the High Energy Astrophysics 
Science Archive Research Center (HEASARC), provided by NASA's Goddard Space 
Flight Center.
This research has made use of the NASA/IPAC Extragalactic Database (\ned)
which is operated by the Jet Propulsion Laboratory, California Institute of
Technology, under contract with NASA. 
This work also made use of the HyperLEDA database
(\href{http://leda.univ-lyon1.fr}{http://leda.univ-lyon1.fr}).
Support for this work was provided by NASA under grant 
NNG04GE76G issued through the Office of Space Sciences Long-Term
Space Astrophysics Program.
\end{acknowledgements}

%\clearpage

\bibliographystyle{apj_hyper}
\bibliography{paper_bibliography.bib}

\end{document}